\documentclass[conference]{IEEEtran}
\IEEEoverridecommandlockouts

\makeatletter
\def\markboth#1#2{\def\leftmark{\@IEEEcompsoconly{\sffamily}\MakeUppercase{\protect#1}}%
\def\rightmark{\@IEEEcompsoconly{\sffamily}\MakeUppercase{\protect#2}}}
\makeatother

\usepackage{graphicx}
\usepackage{epsfig}
\usepackage{epstopdf}
\usepackage{amstext}
\usepackage{amssymb}
\usepackage{amsmath}
\usepackage{amsthm}
\usepackage[font=footnotesize]{caption}
\usepackage{subcaption}
\usepackage{cite}
\usepackage{float}
\usepackage{array}
\usepackage{url}
\usepackage{flexisym}
\usepackage{soul}
\usepackage{color}
\usepackage{bbm}
\usepackage{dsfont}
\usepackage[T1]{fontenc}
\usepackage[utf8]{inputenc}
\usepackage{authblk}
\usepackage[ruled,vlined,linesnumbered]{algorithm2e}

\IEEEoverridecommandlockouts

\begin{document}

\title{Throughput and Coverage for a Mixed Full and Half Duplex Small Cell Network}
%\author{Sanjay Goyal\Mark{1}, Carlo Galiotto\Mark{2}, Nicola Marchetti\Mark{2}, Shivendra Panwar\Mark{1} \\
%\Mark{1}NYU Tandon School of Engineering, Brooklyn, NY, USA \\
%\Mark{2} CTVR, Trinity College, Dublin, Ireland \\
%\{sanjay.goyal,panwar\}@nyu.edu, \{galiotc, marchetn\}@tcd.ie
%}
\author[*]{Sanjay Goyal \thanks {This work is funded by Higher Education Authority under grant HEA/PRTLI Cycle 5 Strand 2 TGI, Science Foundation Ireland through CONNECT grant number 13/RC/2077, NSF award number 1527750, NYSTAR CATT, and NYU Wireless.}}
\author[$\dagger$]{Carlo Galiotto}
\author[$\dagger$]{Nicola Marchetti}
\author[*]{Shivendra Panwar\vspace{-2mm}}
\affil[*]{NYU Tandon School of Engineering, Brooklyn, NY, USA}
\affil[$\dagger$]{CONNECT / The Centre for Future Networks and Communications, Trinity College Dublin, Ireland}
\affil[ ]{{\{sanjay.goyal, panwar\}@nyu.edu, \{galiotc, marchetn\}@tcd.ie}\vspace{-2mm}}
\maketitle

\begin{abstract}
Recent advances in self-interference cancellation enable radios to transmit and receive on the same frequency at the same time. Such a full duplex radio is being considered as a potential candidate for the next generation of wireless networks due to its ability to increase the spectral efficiency of wireless systems. In this paper, the performance of full duplex radio in small cellular systems is analyzed by assuming full duplex capable base stations and half duplex user equipment. However, using only full duplex base stations increases interference leading to outage. We therefore propose a mixed multi-cell system, composed of full duplex and half duplex cells. A stochastic geometry based model of the proposed mixed system is provided, which allows us to derive the outage and area spectral efficiency of such a system. The effect of full duplex cells on the performance of the mixed system is presented under different network parameter settings. We show that the fraction of cells that have full duplex base stations can be used as a design parameter by the network operator to target an optimal tradeoff between area spectral efficiency and outage in a mixed system.
\end{abstract}


%%%%%%%%%%%%%%%%%%%%%%%%%
%%%%%%%%%%%%%%%%%%%%%%%%%

\section{Introduction}\label{sec1}
Recent advances in hardware development \cite{Khandani10,Knox12, Katti13, Duarte13, survey_JSAC} have enabled radios to transmit and receive on the same frequency at the same time, with the potential of doubling the spectral efficiency. Referred to as Full Duplex (FD), these systems are emerging as an attractive solution to the shortage of spectrum for the next generation of wireless networks~\cite{Cisco, NGMN_5G}.

Although FD has the capability of enhancing spectral efficiency, simultaneous downlink and uplink operations on the same band generate additional interference, which is likely to erode the performance gain of FD cells \cite{Goyal_arxiv,Goyal_CommMag}. In this work we focus on a mixed multi-cell system, where only some of the base stations (BSs) operate in FD mode, while the remaining BSs are in half duplex (HD) mode \cite{Goyal_arxiv, Goyal_CommMag, Quekhybrid}.
Using a stochastic geometry-based model that we propose, we investigate the impact of FD cells on the performance of such mixed systems. In particular, we analyze the throughput vs. coverage trade-off of the mixed system as a function of the proportion of FD cells, and for various network parameters such as self-interference cancellation (SIC) levels, and transmit power levels at the BS and at the user equipment (UE).

\subsection{Background and related work}
\begin{figure}[H]
\centering
\includegraphics[width = 2.8in] {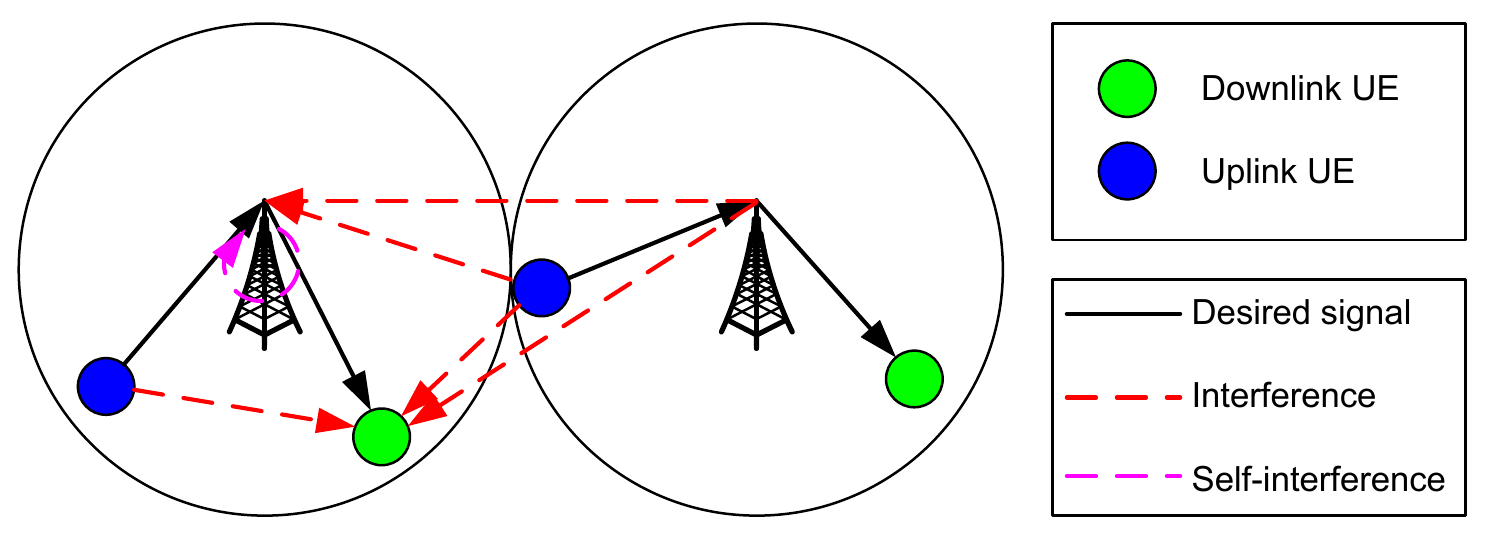}
\vspace{-1mm}
\caption{A full duplex multi-cell interference scenario.}
\label{fig:fig1}
\end{figure}

To successfully achieve SIC, which is required in order to enable FD operation, the FD circuit has a higher cost and power usage. For this reason, it is more practical to implement FD transmission on the infrastructure devices only, whereas the UE  operates in HD mode \cite{Goyal_CommMag}. An example of this is shown in Fig.~\ref{fig:fig1}, where each BS has two UEs scheduled at the same time on the same frequency; one is in uplink, other one is in downlink.
\vspace{-4mm}
\vspace{-2mm}
As we can note from Fig.~\ref{fig:fig1}, in the uplink, a BS receives interference from UEs transmitting in the uplink as well as from BSs of the neighboring cells transmitting in the downlink. It also receives the residual self-interference, generated by the same BS. In the downlink, a UE receives interference from neighboring BSs as well as from all UEs transmitting in the uplink direction. Thus, during FD operation, each direction receives higher interference compared to the HD case. For example, in a HD synchronized system, where in each timeslot all the cells schedule transmission the same direction, the downlink UE receives interference only from the neighboring BSs, and in the uplink the BS  receives interference only from uplink UEs of neighboring cells. As a result of the high interference, FD systems not only cannot achieve their potential  spectral efficiency gain, but can suffer from high outage probability.

Mixed multi-cell systems \cite{Goyal_arxiv, Goyal_CommMag, Quekhybrid}, in which only a given fraction of cells operate in FD mode, have been proposed in order to maintain the interference within a moderate level during FD operations. Although FD cells have the potential of enhancing the area spectral efficiency (ASE) of the network, they also increase the interference, with a consequent drop in terms of coverage.

Among the existing papers addressing FD for wireless networks in multi-cell scenarios, to the best of our knowledge, there is no comprehensive study yet that addresses the ASE vs. coverage trade-off in mixed systems, for both the uplink and the downlink directions.  For instance, some works on stochastic geometry for FD operation in wireless networks have been proposed in \cite{Haenggi_FD,alves2014average, Quekhybrid, elsawy_alpha_duplex}. Tong \emph{et al.}~\cite{Haenggi_FD} investigated the throughput of a wireless network with FD radios using stochastic geometry, but in an ad-hoc setting. Alves \emph{et al.}~\cite{alves2014average} derived the average spectral efficiency for a dense small cell environment and showed the impact of residual self-interference on the performance of FD operation.
Lee \emph{et al.}~\cite{Quekhybrid} derives the throughput of a mixed multi-cell heterogenous network consisting of only downlink and/or FD BSs; however,  this work only focuses on the downlink, while the uplink performance is not considered. An alternative approach to a multi-cell network with FD operation in each cell is considered in \cite{elsawy_alpha_duplex}, where the authors proposed a scheme which allows a partial overlap between uplink and downlink frequency bands. In \cite{elsawy_alpha_duplex}, it is shown that the amount of the overlap can be optimized to achieve the maximum FD gain. However, all the papers mentioned above: \cite{alves2014average, Quekhybrid, elsawy_alpha_duplex} assume the UEs to have FD capabilities, which is neither practical nor economical, given existing FD circuit designs \cite{survey_JSAC}. Moreover, most of the existing work investigates the ASE, while increase in outage probability is not taken into account as a metric to assess the system.

\subsection{Contribution}
In this paper, we consider a mixed multi-cell system, in which BSs can operate either in FD or in HD mode, while UEs only operate in HD mode. The main contributions of our work are, (i) we propose a model based on stochastic geometry that allows us to characterize the outage probability and the ASE of both BSs and UEs, for both FD and HD cells; (ii) we investigate the  ASE vs. coverage trade-off of mixed systems for different network parameters and we aim at finding the proportion of FD BSs such that some given constraints in terms of ASE or, alternatively, of coverage, can be met.

Among our main findings, we show that the fraction of FD cells can be used as a design parameter to target different ASE vs. coverage trade-offs for the network operator; in particular, by increasing the amount of  FD cells in the mixed system, the overall throughput increases at the cost of a drop in terms coverage, and vice-versa.

%%%%%%%%%%%%%%%%%%%%%%%%%
\vspace{-1mm}
%%%%%%%%%%%%%%%%%%%%%%%%%

\section{System Model}\label{sec:Model}
We consider a network of small cell BSs deployed according to a homogeneous and isotropic Spatial Poisson Point Process (SPPP) $\Phi_{\mathrm{B}}$ with density $\lambda_{\mathrm{B}}$. We also consider
a set of UEs, whose locations can be modeled as a SPPP $\Phi_{\mathrm{U}}$ with density $\lambda_{\mathrm{U}}$, where $\lambda_{\mathrm{U}} >> \lambda_{\mathrm{B}}$. The BSs are assumed to be capable of FD operation, while the UEs are limited to HD operation. We focus on a single LTE subframe, where all cells are assumed be synchronized in terms of subframe alignment. At any subframe in any cell the BS can either be in FD mode or HD mode. In the case of HD mode the transmission can be either in downlink direction or in uplink direction.

We assume that each UE, either in the uplink or in the downlink, is served by the nearest BS. This deployment of BSs create a Voronoi tessellation with several cells, each represented by the Voronoi region around the BS location. We further assume that each FD BS will have one uplink UE and one downlink UE scheduled simultaneously at the same subframe whereas each downlink HD BS will have one downlink UE and each uplink HD BS will have one uplink UE active at their subframe. 

We define $\rho_{\mathrm{F}}$, $\rho_{\mathrm{D}}$, and $\rho_{\mathrm{U}}$ as the probability of a BS to be in FD mode, downlink HD mode, and uplink HD mode, respectively, with $\rho_{\mathrm{F}}+\rho_{\mathrm{D}}+\rho_{\mathrm{U}}=1$. From the ``Thinning theorem'' \cite[Sec 2.36]{haenggi2013stochastic}, the locations of the FD BSs, of the downlink HD BSs, and of the uplink HD BSs follow the SPPP, which we denote as $\Phi_{\mathrm{B}}^{\mathrm{F}}$, $\Phi_{\mathrm{B}}^{\mathrm{D}}$, and $\Phi_{\mathrm{B}}^{\mathrm{U}}$ and have densities $\rho_{\mathrm{F}} \lambda_{\mathrm{B}} $, $\rho_{\mathrm{D}} \lambda_{\mathrm{B}}$, and  $\rho_{\mathrm{U}} \lambda_{\mathrm{B}}$, respectively; furthermore, we assume the processes $\Phi_{\mathrm{B}}^{\mathrm{F}}$, $\Phi_{\mathrm{B}}^{\mathrm{D}}$, and $\Phi_{\mathrm{B}}^{\mathrm{U}}$ to be independent of one another.  

Because of the assumptions of our system model, at most two UEs per cell are active, one in HD cells and two in FD cells. In other words, among all the UEs in the network, we focus our analysis only on narrow subsets of $\Phi_{\mathrm{U}}$; specifically, we consider: (i) the set of downlink and uplink UEs served by the FD BSs, namely $\tilde{\Phi}_{\mathrm{U}}^{\mathrm{F,D}}$ and $\tilde{\Phi}_{\mathrm{U}}^{\mathrm{F,U}}$, respectively; (ii) the subset of downlink and uplink UEs served by the HD BSs, namely $\tilde{\Phi}_{\mathrm{U}}^{\mathrm{H,D}}$ and $\tilde{\Phi}_{\mathrm{U}}^{\mathrm{H,U}}$, respectively. Due to the association strategy used to assign the UEs to a given cell $\tilde{\Phi}_{\mathrm{U}}^{\mathrm{F,D}}$, $\tilde{\Phi}_{\mathrm{U}}^{\mathrm{F,U}}$, $\tilde{\Phi}_{\mathrm{U}}^{\mathrm{H,D}}$, and $\tilde{\Phi}_{\mathrm{U}}^{\mathrm{H,U}}$ are not SPPPs. Nonetheless, to maintain the mathematical tractability, we approximate these subsets as SPPP; this has been proved a good approximation in \cite{jeffAndrews_uplink,dynamic_TDD}. From the ``Thinning theorem'' \cite[Sec 2.36]{haenggi2013stochastic}, we can obtain the densities of these subsets. To summarize, below we report the definitions
of the active UEs' subsets with the related densities.

\begin{itemize}
\item $\tilde{\Phi}_{\mathrm{U}}^{\mathrm{F,D}} \subset \Phi_{\mathrm{U}}$, with density $\lambda_{\mathrm{U,F,D}} = \rho_{\mathrm{F}}\lambda_{\mathrm{B}}$, is the subset of active downlink UEs served by the FD BSs;
\item $\tilde{\Phi}_{\mathrm{U}}^{\mathrm{F,U}} \subset \Phi_{\mathrm{U}}$, with density $\lambda_{\mathrm{U,F,U}} = \rho_{\mathrm{F}}\lambda_{\mathrm{B}}$, is the subset of active uplink UEs served by the FD BSs;
\item $\tilde{\Phi}_{\mathrm{U}}^{\mathrm{H,D}} \subset \Phi_{\mathrm{U}}$, with density $\lambda_{\mathrm{U,H,D}} = \rho_{\mathrm{D}}\lambda_{\mathrm{B}}$, is the subset of active downlink UEs served by the HD BSs;
\item $\tilde{\Phi}_{\mathrm{U}}^{\mathrm{H,U}} \subset \Phi_{\mathrm{U}}$, with density $\lambda_{\mathrm{U,H,U}} = \rho_{\mathrm{U}}\lambda_{\mathrm{B}}$, is the subset of active uplink UEs served by the HD BSs.
\end{itemize}

For ease of notation, we consider the set $\tilde{\Phi}_{\mathrm{U}}$ of all active UEs, which is the union $\tilde{\Phi}_{\mathrm{U}}^{\mathrm{F,D}} \cup \tilde{\Phi}_{\mathrm{U}}^{\mathrm{F,U}} \cup \tilde{\Phi}_{\mathrm{U}}^{\mathrm{H,D}} \cup \tilde{\Phi}_{\mathrm{U}}^{\mathrm{H,U}}$; $\tilde{\Phi}_{\mathrm{U}}$ is assumed to be an SPPP and its density is the sum of each subset's density, which is~($\rho_{\mathrm{F}}$ + 1)$\lambda_{\mathrm{B}}$~\cite[Preposition 1.3.3]{baccellistochastic}.

Note that, the set of interfering UEs and BSs would be correlated due to the association technique mentioned above. However, to maintain  model tractability, we assume that the set of interfering UEs is independent of the set of interfering BSs; this assumption has been proved to provide a good approximation for the results in previous works \cite{alves2014average, elsawy_alpha_duplex}. Moreover, we
also assume the SPPPs $\tilde{\Phi}_{\mathrm{U}}^{\mathrm{F,D}}$, $\tilde{\Phi}_{\mathrm{U}}^{\mathrm{F,U}}$, $\tilde{\Phi}_{\mathrm{U}}^{\mathrm{H,D}}$, and $\tilde{\Phi}_{\mathrm{U}}^{\mathrm{H,U}}$ to be independent of one another and independent of $\Phi_{\mathrm{B}}^{\mathrm{F}}$, $\Phi_{\mathrm{B}}^{\mathrm{D}}$, and $\Phi_{\mathrm{B}}^{\mathrm{U}}$.

\vspace{-1mm}
%%%%%%%%%%%%%%%%%%%%%%%%%
\subsection{Channel model}\label{sec:ChModel}
In our analysis, we model the different links with different parameters. In general, BSs and UEs are different kinds of nodes in terms of antenna height, antenna characteristics, mobility, etc. For example, different channel models are recommended by 3GPP for BS-to-BS, BS-to-UE, and UE-to-UE links \cite{3GPP:1}. We considered the following path loss models for the different links that exist in our system:
\begin{itemize}
\item BS-to-UE path loss $\mathrm{PL}_{1}(d)=K_{1}d^{-\alpha_{1}}$.
\item UE-to-BS path loss $\mathrm{PL}_{1}(d)=K_{1}d^{-\alpha_{1}}$.
\item UE-to-UE path loss $\mathrm{PL}_{2}(d)=K_{2}d^{-\alpha_{2}}$.
\item BS-to-BS path loss $\mathrm{PL}_{3}(d)=K_{3}d^{-\alpha_{3}}$.
\end{itemize}
where $\alpha_{1}$, $\alpha_{2}$, and $\alpha_{3}$ are the path loss exponents; $K_{1}$, $K_{2}$, and $K_{3}$ are the signal attenuations at distance $d$ = 1. We further assume that the propagation is affected by Rayleigh fading, which is exponentially distributed $\sim$ exp($\mu$) with mean $\mu^{-1}$. In the next subsections, we use $g$, $h$, $g'$, and $h'$ to denote Rayleigh fading for the BS-to-UE link, UE-to-UE link, BS-to-BS link, and UE-to-BS link, respectively.

%%%%%%%%%%%%%%%%%%%%%%%%%
\vspace{-1mm}
\subsection{BS and UE Transmit Power Allocation}\label{sec:Powalloc}
We model downlink transmission with a fixed power transmission scheme. All the BSs transmit with power $P_{\mathrm{B}}$. For uplink modeling, we use distance-proportional fractional power control \cite{jeffAndrews_uplink}, in which each UE, which is at distance $R$ from its serving BS transmits with power  $ P_{\mathrm{U}} K_{1}^{-\epsilon }R^{\epsilon \alpha_{1}}$, where $\epsilon       \in [0,1]$ is the power control factor. If $\epsilon = 1$, the path loss is completely compensated, and if $\epsilon = 0$ all UEs transmit with the same power $P_{\mathrm{U}}$. Both antennas at the BS and at the UE are assumed to be isotropic.

%%%%%%%%%%%%%%%%%%%%%%%%%
\vspace{-1mm}
%%%%%%%%%%%%%%%%%%%%%%%%%
\section{SINR Distributions}\label{sec:MainSINR}
In this section we present the analytic results for signal to interference and noise ratio (SINR) distributions in our mixed system for both downlink and uplink. We are interested in evaluating the SINR Complementary Cumulative Distribution Function (CCDF), which can written as

\begin{equation}\label{eq:sinr_ccdf_general}
 \mathrm{P}[\gamma > y] = \mathbb{E}_r [\mathrm{P}[\gamma > y|r]] = \int_0^{\infty} \mathrm{P}[\gamma > y|r=R] f_r(R) \mathrm{d}R,
\end{equation}
where $\gamma$ denotes the SINR, $\mathbb{E}_r$ denotes the expectation over $r$, $f_r(R)$ is the Probability Density Function (PDF) of the distance $r$ of the receiver of interest to the transmitter. 

We focus the analysis on the typical receiver, either the UE or the BS, depending on whether we consider the downlink or the uplink, respectively. We differentiate the SINR expression for the downlink and the uplink in Section~\ref{sec:FDdownlinkSINR} and~\ref{sec:FDuplinkSINR}, respectively.

\subsection{Probability Density Function of the Distance to the Closest Transmitter}\label{sec:Th_ANA_pdf_dis}
It is known from the literature that the distance $r$ of a given point to the closest point of an SPPP with density $\lambda$ has the following PDF~\cite{haenggi2013stochastic, Andrews_classic}:
\begin{equation}\label{eq:distance_dis_pdf_gen}
f_r(R) = e^{-\pi \lambda R^2} 2 \pi \lambda R.
\end{equation}

We recall from Section~\ref{sec:Model} that in our system model we assume two sets, one for the set of BSs (i.e., $\Phi_{\mathrm{B}}$), the other for the set of active UEs (i.e., $\tilde{\Phi}_{\mathrm{U}}$).
Although we assume the independence between $\Phi_{\mathrm{B}}$ and $\tilde{\Phi}_{\mathrm{U}}$, some correlation exists between them. Moreover, $\tilde{\Phi}_{\mathrm{U}}$ would not be an SPPP,
despite we assume it to be as such to maintain the mathematical tractability. Because of this, equation (\ref{eq:distance_dis_pdf_gen}), which models the PDF for SPPPs, might not model accurately the PDF of 
the distance from a UE (i.e, any point of $\tilde{\Phi}_{\mathrm{U}}$) to the closest BS (i.e. a point of $\Phi_{\mathrm{B}}$). A solution to this problem has been proposed by the authors in~\cite{dynamic_TDD}, 
who suggested to replace~(\ref{eq:distance_dis_pdf_gen}) with the following function:
\begin{equation}\label{eq:distance_dis_pdf_mod}
f_r(R) = e^{-\pi \lambda_{\mathrm{B}} R^2} 2 \pi \nu \lambda_{\mathrm{B}} R,
\end{equation}
where $\nu$ is a correction factor that takes into account the effect of the correlation among points on the distance distribution; specifically, the authors of~\cite{dynamic_TDD} have proposed to use 1.25 as a value of $\nu$. One should note that~(\ref{eq:distance_dis_pdf_gen}) can be obtained as a special case of~(\ref{eq:distance_dis_pdf_mod}) for $\nu = 1$. The CDF corresponding to~(\ref{eq:distance_dis_pdf_mod}) is the following: 
\begin{equation}\label{eq:distance_dis_cdf_mod}
\mathbb{P}\{ r \leq R\} = 1-\mathrm{exp}(-\pi \nu \lambda_{\mathrm{B}} R^2), R \geq 0.
\vspace{-1mm}
\end{equation}

In our work, we will make use of simulation results to determine which function between~(\ref{eq:distance_dis_pdf_gen}) and~(\ref{eq:distance_dis_pdf_mod}) provides the best match with the analytical results. We will show this in~Appendix~\ref{app:A}.

%%%%%%%%%%%%%%%%%%%%%%%%%
\subsection{Downlink SINR in a FD Cell of the Mixed System}\label{sec:FDdownlinkSINR}
The SINR at a downlink UE of interest in a FD cell of the mixed system, is given by
\vspace{-2mm}
\begin{equation}\label{eq:sinr_fd_dwn}
\gamma_{\mathrm{FD,UE}}=\frac{P_{\mathrm{RX,UE}}}{N_{0}+I_{\mathrm{D}}+I_{\mathrm{U}}},
\vspace{-1mm}
\end{equation}
where $N_{0}$ is the thermal noise power at the downlink UE, and $P_{\mathrm{RX,UE}}$ is the received signal power at the downlink UE from its serving BS, which is given by
\vspace{-1mm}
\begin{equation}\label{eq:p_rx}
P_{\mathrm{RX,UE}} = P_{\mathrm{B}} g_{b_{0}} K_{1}r^{-\alpha_{1}},
\vspace{-1mm}
\end{equation}
where $r$ is the distance between the downlink UE and its serving BS. The serving BS is indicated by $b_0$, and $g_{b_{0}}$ denotes the Rayleigh fading affecting the signal from the BS $b_0$. $I_{\mathrm{D}}$ and $I_{\mathrm{U}}$ are the total interference received at the downlink UE from all the downlink transmissions, and from all the uplink transmissions, respectively. The total interference from all the downlink transmissions including all FD cells ($\Phi_{\mathrm{B}}^{\mathrm{F}} \backslash{b_0}$) and all HD downlink cells ($\Phi_{\mathrm{B}}^{\mathrm{D}}$) can be defined as
\vspace{-1mm}
\begin{equation}\label{eq:int_D}
I_{\mathrm{D}} = P_{\mathrm{B}} \sum_{b      \in \{\Phi_{\mathrm{B}}^{\mathrm{D}}       \cup       \Phi_{\mathrm{B}}^{\mathrm{F}} \backslash{b_0}\}}\: {g_b K_{1}R_b^{-\alpha_{1}}},
\vspace{-1mm}
\end{equation}
where $R_b$ is the distance of the downlink UE from the neighboring active BS $b$, and $g_{b}$ denotes the Rayleigh fading for this link. Similarly, $I_{\mathrm{U}}$ is the sum of interference from the uplink transmission of FD cells and HD uplink cells,
\vspace{-1mm}
\begin{equation}\label{eq:int_U}
I_{\mathrm{U}} = P_{\mathrm{U}} \sum_{u      \in \{\tilde{\Phi}_{\mathrm{U}}^{\mathrm{F,U}}       \cup       \tilde{\Phi}_{\mathrm{U}}^{\mathrm{H,U}} \}}\: K_{1}^{-\epsilon }Z_{u}^{\epsilon \alpha_1} {h_u K_{2}D_u^{-\alpha_{2}}},
\vspace{-1mm}
\end{equation}
where the general uplink UE $u$, (i) is located at distance $Z_{u}$ from its serving BS; (ii) transmits with power $P_{\mathrm{U}} K_{1}^{-\epsilon }Z_{u}^{\epsilon \alpha_1}$; (iii) is at distance $D_u$ from the downlink UE of interest. The symbol $h_u$ denotes the Rayleigh fading for the channel between uplink UE $u$ and the downlink UE of interest.

%%%%%%%%%%%%%%%%%%%%%%%%%
\subsubsection{Downlink SINR CCDF}\label{sec:FDdownlinkSINRCCDF}

By using (\ref{eq:sinr_fd_dwn}) and (\ref{eq:p_rx}), we can define,
\begin{equation*}
\begin{aligned}
& \mathrm{P}[\gamma_{\mathrm{FD,UE}} > y|r=R] =  \mathrm{P} \left[ \frac{P_{\mathrm{B}} g_{b_{0}} K_{1}R^{-\alpha_{1}}}{N_{0}+I_{\mathrm{D}}+I_{\mathrm{U}}}  > y \right] \\
& = \mathrm{P} \left[ g_{b_{0}} > y P_{\mathrm{B}}^{-1} K_{1}^{-1} R^{\alpha_{1}} (N_{0}+I_{\mathrm{D}}+I_{\mathrm{U}}) \right] \\
\end{aligned}
\end{equation*}
\begin{equation}\label{eq:sinr_ccdf_2}
\overset{(a)} = e^{-\mu y P_{\mathrm{B}}^{-1} K_{1}^{-1} R^{\alpha_{1}} N_{0}}~\mathcal{L}_{I_{\mathrm{D}} + I_{\mathrm{U}}}(\mu y P_{\mathrm{B}}^{-1} K_{1}^{-1} R^{\alpha_{1}}),
\end{equation}
%\begin{equation}\label{eq:sinr_ccdf_2}
%\begin{split}
%& \mathrm{P}[\gamma_{\mathrm{FD,UE}} > y|r=R] = \mathrm{P} \left[ \frac{P_{\mathrm{B}} g_{b_{0}} K_{1}R^{-\alpha_{1}}}{N_{0}+I_{\mathrm{D}}+I_{\mathrm{U}}}  > y \right] \\
%& = \mathrm{P} \left[ g_{b_{0}} > y P_{\mathrm{B}}^{-1} K_{1}^{-1} R^{\alpha_{1}} (N_{0}+I_{\mathrm{D}}+I_{\mathrm{U}}) \right] \\
%&  \overset{(a)} = e^{-\mu y P_{\mathrm{B}}^{-1} K_{1}^{-1} R^{\alpha_{1}} N_{0}}~\mathcal{L}_{I_{\mathrm{D}} + I_{\mathrm{U}}}(\mu y P_{\mathrm{B}}^{-1} K_{1}^{-1} R^{\alpha_{1}}),
%\end{split}
%\end{equation}
where (a) follows from the fact that $g_{b_{0}}$ $\sim$ $\exp(\mu )$. The Laplace transform of the total interference $(I_{\mathrm{D}} + I_{\mathrm{U}})$, $\mathcal{L}_{I_{\mathrm{D}} + I_{\mathrm{U}}}(s)$, where $s = \mu y P_{\mathrm{B}}^{-1} K_{1}^{-1} R^{\alpha_{1}}$, can be written as
\begin{equation}\label{eq:laplace_FD_D_1}
\begin{aligned}
 &\mathcal{L}_{I_{\mathrm{D}} + I_{\mathrm{U}}}(s) = \\
 &\mathbb{E}_{\Phi_{\mathrm{B}}^{\mathrm{F}}       \cup       \Phi_{\mathrm{B}}^{\mathrm{D}}       \cup      \tilde{ \Phi}_{\mathrm{U}}^{\mathrm{F,U}} \cup      \tilde{ \Phi}_{\mathrm{U}}^{\mathrm{H,U}}, g_b, h_u, Z_{u}}  \Bigg[ e^{-s \sum_{b      \in \{\Phi_{\mathrm{B}}^{\mathrm{D}}       \cup       \Phi_{\mathrm{B}}^{\mathrm{F}} \backslash{b_0}\}}\: {g_b P_{\mathrm{B}} K_{1}R_b^{-\alpha_{1}}}} \\
&     \times    e^{-s \sum_{u      \in \{\tilde{\Phi}_{\mathrm{U}}^{\mathrm{F,U}}       \cup      \tilde{\Phi}_{\mathrm{U}}^{\mathrm{H,U}} \}}\: {h_u P_{\mathrm{U}} K_1^{-\epsilon } Z_u^{\epsilon \alpha_1} K_{2}D_u^{-\alpha_{2}}})} \Bigg].
\end{aligned}
\vspace{-1mm}
\end{equation}
%\begin{equation}\label{eq:laplace_FD_D_1}
% \mathcal{L}_{I_{\mathrm{D}} + I_{\mathrm{U}}}(s) = \mathbb{E}_{\Phi_{\mathrm{B}}^{\mathrm{F}}       \cup       \Phi_{\mathrm{B}}^{\mathrm{D}}       \cup       \Phi_{\mathrm{B}}^{\mathrm{U}},~g_b,h_u} \left[ \mathrm{exp} \left( -s \sum_{b      \in \{\Phi_{\mathrm{B}}^{\mathrm{D}}       \cup       \Phi_{\mathrm{B}}^{\mathrm{F}} \backslash{b_0}\}}\: {g_b P_{\mathrm{B}} K_{1}R_b^{-\alpha_{1}}} \right)~\mathrm{exp} \left( -s \sum_{u      \in \{\Phi_{\mathrm{B}}^{\mathrm{U}}       \cup       \Phi_{\mathrm{B}}^{\mathrm{F}} \}}\: {h_u P_{\mathrm{U}} K_{2}D_u^{-\alpha_{2}}})\right) \right]
%\end{equation}

Using the independence among $\Phi_{\mathrm{B}}^{\mathrm{F}}$, $\Phi_{\mathrm{B}}^{\mathrm{D}}$, $\tilde{\Phi}_{\mathrm{U}}^{\mathrm{F,U}}$, and $\tilde{\Phi}_{\mathrm{U}}^{\mathrm{H,U}}$ mentioned in Section~\ref{sec:Model}, we can separate the expectation to obtain:
\begin{equation}\label{eq:laplace_FD_D_2}
\begin{aligned}
 &\mathcal{L}_{I_{\mathrm{D}} + I_{\mathrm{U}}}(s) = \underbrace{\mathbb{E}_{\Phi_{\mathrm{B}}^{\mathrm{F}}       \cup       \Phi_{\mathrm{B}}^{\mathrm{D}}, g_b} \Bigg[ e^{-s \sum_{b      \in \{\Phi_{\mathrm{B}}^{\mathrm{D}}       \cup       \Phi_{\mathrm{B}}^{\mathrm{F}} \backslash{b_0}\}}\: {g_b P_{\mathrm{B}} K_{1}R_b^{-\alpha_{1}}}}\Bigg]}_{L_x(s)}      \\
&  \times  \underbrace{\mathbb{E}_{\tilde{\Phi}_{\mathrm{U}}^{\mathrm{F,U}}      \cup       \tilde{\Phi}_{\mathrm{U}}^{\mathrm{H,U}}, h_u, Z_u}\Bigg[e^{-s \sum_{u      \in \{\tilde{\Phi}_{\mathrm{U}}^{\mathrm{F,U}}      \cup       \tilde{\Phi}_{\mathrm{U}}^{\mathrm{H,U}} \}}\: {\frac{h_u P_{\mathrm{U}} Z_u^{\epsilon \alpha_1}  K_{2}}{K_1^{\epsilon } D_u^{\alpha_{2}}} })} \Bigg]}_{L_y(s)}.
\end{aligned}
\vspace{-1mm}
\end{equation}

The first term can be further written as
\vspace{-0mm}
\begin{equation}\label{eq:laplace_FD_D_3}
\begin{aligned}
 L_x(s) = \mathbb{E}_{\Phi_{\mathrm{B}}^{\mathrm{F}}, g_b} & \Bigg[ e^{ -s \sum_{b      \in \Phi_{\mathrm{B}}^{\mathrm{F}} \backslash{b_0}}\: {g_b P_{\mathrm{B}} K_{1}R_b^{-\alpha_{1}}} }\Bigg]  \\
&    \times    \mathbb{E}_{\Phi_{\mathrm{B}}^{\mathrm{D}}, g_b} \Bigg[ e^{ -s \sum_{b      \in \Phi_{\mathrm{B}}^{\mathrm{D}}}\: {g_b P_{\mathrm{B}} K_{1}R_b^{-\alpha_{1}}} }\Bigg].
\end{aligned}
\vspace{-0mm}
\end{equation}

By applying the Probability Generating Functional (PGFL) \cite{haenggi2013stochastic} of the SPPP to (\ref{eq:laplace_FD_D_3}), it can be further written as:
%\begin{equation}\label{eq:laplace_FD_D_4}
% \mathrm{exp} \left( -2\pi \rho_{\mathrm{F}} \lambda_{\mathrm{B}} \int_{R}^{\infty} \left(\frac{s K_{1} P_{\mathrm{B}} v^{-\alpha_{1}}}{s K_{1} P_{\mathrm{B}} v^{-\alpha_{1}} + \mu } \right) v \mathrm{d}v \right)       \cdot \mathrm{exp} \left( -2\pi \rho_{\mathrm{D}} \lambda_{\mathrm{B}} \int_{R}^{\infty} \left(\frac{s K_{1} P_{\mathrm{B}} v^{-\alpha_{1}}}{s K_{1} P_{\mathrm{B}} v^{-\alpha_{1}} + \mu } \right) v \mathrm{d}v \right).
%\end{equation}
%
%We can simplify \eqref{eq:laplace_FD_D_4} as follows:
\vspace{-1mm}
\begin{equation}\label{eq:laplace_FD_D_4_A}
L_x(s) = e^ { -2\pi \lambda_{\mathrm{B}}( \rho_{\mathrm{F}} + \rho_{\mathrm{D}} ) \int_{R}^{\infty} \left(\frac{s K_{1} P_{\mathrm{B}} v^{-\alpha_{1}}}{s K_{1} P_{\mathrm{B}} v^{-\alpha_{1}} + \mu } \right) v \mathrm{d}v }.
\vspace{-1mm}
\end{equation}

Similarly the second term in (\ref{eq:laplace_FD_D_2}) can be written as:
%\begin{equation}\label{eq:laplace_FD_D_5}
%\mathrm{exp} \left( -2\pi \rho_{\mathrm{F}} \lambda_{\mathrm{B}} \int_{0}^{\infty} \left(\frac{s K_{2} P_{\mathrm{U}} v^{-\alpha_{2}}}{s K_{2} P_{\mathrm{U}} v^{-\alpha_{2}} + \mu } \right) v \mathrm{d}v \right)       \cdot \mathrm{exp} \left( -2\pi \rho_{\mathrm{U}} \lambda_{\mathrm{B}} \int_{0}^{\infty} \left(\frac{s K_{2} P_{\mathrm{U}} v^{-\alpha_{2}}}{s K_{2} P_{\mathrm{U}} v^{-\alpha_{2}} + \mu } \right) v \mathrm{d}v \right).
%\end{equation}

%\begin{equation}\label{eq:laplace_FD_D_5}
%\begin{aligned}
% \mathrm{exp} &\left( -2\pi \rho_{\mathrm{F}} \lambda_{\mathrm{B}} \int_{0}^{\infty}  \left(1 - \mathbb{E}_{Z_u}\left[\frac{\mu }{s K_{2} P_{\mathrm{U}} K_1^{-\epsilon } Z_u^{\epsilon \alpha_1} v^{-\alpha_{2}} + \mu }\right] \right) v \mathrm{d}v \right)      \cdot \\
%& \mathrm{exp} \left( -2\pi \rho_{\mathrm{U}} \lambda_{\mathrm{B}} \int_{0}^{\infty} \left(1-\mathbb{E}_{Z_u}\left[\frac{\mu }{s K_{2} P_{\mathrm{U}} K_1^{-\epsilon } Z_u^{\epsilon \alpha_1} v^{-\alpha_{2}} + \mu } \right] \right) v \mathrm{d}v \right).
%\end{aligned}
%\end{equation}

\vspace{-4mm}
\begin{equation}\label{eq:laplace_FD_D_5}
\begin{aligned}
&L_y(s) =  \\
& e^{ -2\pi (\rho_{\mathrm{F}} + \rho_{\mathrm{U}}) \lambda_{\mathrm{B}} \int_{0}^{\infty}  \Bigg(1 - \mathbb{E}_{Z_u}\left[\frac{\mu }{s K_{2} P_{\mathrm{U}} K_1^{-\epsilon } Z_u^{\epsilon \alpha_1} v^{-\alpha_{2}} + \mu }\right] \Bigg) v \mathrm{d}v} .
\end{aligned}
\vspace{-1mm}
\end{equation}

Note that in (\ref{eq:laplace_FD_D_4_A}), the lower extreme of integration is $R$ because the closest interferer BS (either FD or HD) from the FD downlink UE of interest is at least at a distance $R$. However, the closest uplink UE interferer of a FD cell can also be in its own cell, so the lower extreme of integration in (\ref{eq:laplace_FD_D_5}) is zero.
Under the special case of no power control {$\epsilon = 0$}, expression (\ref{eq:laplace_FD_D_5}) is converted to:
%\begin{equation}\label{eq:laplace_FD_D_6}
%\mathrm{exp} \left( -2\pi \rho_{\mathrm{F}} \lambda_{\mathrm{B}} \int_{0}^{\infty} \left(\frac{s K_{2} P_{\mathrm{U}} v^{-\alpha_{2}}}{s K_{2} P_{\mathrm{U}} v^{-\alpha_{2}} + \mu } \right) v \mathrm{d}v \right)       \cdot \mathrm{exp} \left( -2\pi \rho_{\mathrm{U}} \lambda_{\mathrm{B}} \int_{0}^{\infty} \left(\frac{s K_{2} P_{\mathrm{U}} v^{-\alpha_{2}}}{s K_{2} P_{\mathrm{U}} v^{-\alpha_{2}} + \mu } \right) v \mathrm{d}v \right).
%\end{equation}
%
%Still, we can simplify \eqref{eq:laplace_FD_D_6_A} as follows:
\vspace{-1mm}
\begin{equation}\label{eq:laplace_FD_D_6_A}
L_y(s) = e^{-2\pi \lambda_{\mathrm{B}} (\rho_{\mathrm{F}} + \rho_{\mathrm{U}}) \int_{0}^{\infty} \left(\frac{s K_{2} P_{\mathrm{U}} v^{-\alpha_{2}}}{s K_{2} P_{\mathrm{U}} v^{-\alpha_{2}} + \mu } \right) v \mathrm{d}v }.
\end{equation}

Finally, by plugging~(\ref{eq:sinr_ccdf_2}) in~(\ref{eq:sinr_ccdf_general}), we obtain the CCDF of the downlink SINR in a FD cell for a mixed system.

\begin{equation}\label{eq:sinr_downlink_finally}
\begin{split}
 &\mathrm{P}[\gamma_{FD,BS} > y] = \int_0^{\infty} \mathrm{P}[\gamma_{FD,BS} > y|r=R] f_r(R) \mathrm{d}R = \\
 & \int_0^{\infty} e^{-\mu y P_{\mathrm{B}}^{-1} K_{1}^{-1} R^{\alpha_{1}} N_{0}}~\mathcal{L}_{x}(s)~\mathcal{L}_{y}(s) f_r(R) \mathrm{d}R,
 \end{split}
\end{equation}
where $s = \mu y P_{\mathrm{B}}^{-1} K_{1}^{-1} R^{\alpha_{1}}$, and $f_r(R)$ is given by~(\ref{eq:distance_dis_pdf_mod}); note that we will determine the value to be used for the parameter $\nu$ of~(\ref{eq:distance_dis_pdf_mod}) in Appendix~\ref{app:A}, where we will compare the analytical model with simulation results.
%%%%%%%%%%%%%%%%%%%%%%%%%
\subsection{Uplink SINR in a FD Cell of the Mixed System}\label{sec:FDuplinkSINR}
The SINR for the uplink UE of interest in a FD cell of the mixed system, is given by
\vspace{-2mm}
\begin{equation}\label{eq:sinr_fd_up}
\gamma_{\mathrm{FD,BS}}=\frac{P_{\mathrm{RX,BS}}}{N_{1}+I'_{\mathrm{D}}+I'_{\mathrm{U}}+ \mathcal{C}(P_{\mathrm{B}})},
\vspace{-1mm}
\end{equation}
where $P_{\mathrm{RX,BS}}$ is the received signal power from the uplink UE of interest to its serving BS, which is given by
\vspace{-1mm}
\begin{equation}\label{eq:p_rx_up}
P_{\mathrm{RX,BS}} = P_{\mathrm{U}} h'_{u_{0}} K_{1}^{(1-\epsilon )}r^{\alpha_{1}(\epsilon-1)},
\vspace{-1mm}
\end{equation}
where $r$ is the distance between the uplink UE and its serving BS, and $h'_{u_{0}}$ denotes the Rayleigh fading for this link. In (\ref{eq:sinr_fd_up}), $N_1$ is the thermal noise power at the BS receiver and $\mathcal{C}(P_{\mathrm{B}})$ is the residual self-interference at the BS, which depends on the transmit power of the BS, $P_{\mathrm{B}}$. We model the residual self-interference as Gaussian noise, the power of which equals the ratio of the transmit power of the BS, $P_{\mathrm{B}}$, and the amount of SIC~\cite{Goyal_CommMag}.

In (\ref{eq:sinr_fd_up}), $I'_{\mathrm{D}}$ and $I'_{\mathrm{U}}$ are the total interference received at the BS from all the downlink transmissions, and from all the uplink transmissions, respectively. These can be defined as
\begin{equation}\label{eq:int_D_up}
I'_{\mathrm{D}} = P_{\mathrm{B}} \sum_{b      \in \{\Phi_{\mathrm{B}}^{\mathrm{D}}       \cup       \Phi_{\mathrm{B}}^{\mathrm{F}} \backslash{b_0}\}}\: {g'_b K_{3}L_b^{-\alpha_{3}}},
\end{equation}

\begin{equation}\label{eq:int_U_up}
I'_{\mathrm{U}} = P_{\mathrm{U}} \sum_{u      \in \{\tilde{\Phi}_{\mathrm{U}}^{\mathrm{F,U}}       \cup       \tilde{\Phi}_{\mathrm{U}}^{\mathrm{H,U}} \}: X_u > Z_u }\: {h'_u K_{1}^{(1-\epsilon )} Z_u^{\epsilon \alpha_1}X_u^{-\alpha_{1}}},
\end{equation}
where $L_b$ and $X_u$ are the distances of the BS from its neighboring BS $b$ and the active uplink UE $u$ in a neighboring cell, respectively; $Z_u$ is the distance of the UE $u$ from its serving
BS. As proposed in~\cite{dynamic_TDD}, the condition $\{X_u > Z_u \}$ in~(\ref{eq:int_U_up}) for all $u      \in \{\Phi_{\mathrm{B}}^{\mathrm{U}}       \cup       \Phi_{\mathrm{B}}^{\mathrm{F}} \}$ guarantees that the distance $Z_u$ of the interfering UE $u$ to its serving BS is shorter than the distance from $u$ to the victim BS.

%%%%%%%%%%%%%%%%%%%%%%%%%
\subsubsection{Uplink SINR CCDF}\label{sec:FDuplinkSINRCCDF}

The CCDF for the uplink SINR, $\gamma_{\mathrm{FD,BS}}$, is given by,
\begin{equation}\label{eq:sinr_ccdf_u_1}
\mathrm{P}[\gamma_{\mathrm{FD,BS}} > y] = \int_0^{\infty} \mathrm{P}[\gamma_{\mathrm{FD,BS}} > y|r=R] f_r(R) \mathrm{d}R.
\end{equation}

By using the similar steps described in Section~\ref{sec:FDdownlinkSINRCCDF},
\vspace{0mm}
\begin{equation}\label{eq:sinr_ccdf_u_2}
\begin{aligned}
\mathrm{P}[\gamma_{\mathrm{FD,BS}} > y|r=R] =~&e^{-\mu y P_{\mathrm{U}}^{-1} K_{1}^{(\epsilon-1)} R^{\alpha_{1}(1-\epsilon )} (N_{1}+\mathcal{C}(P_{\mathrm{B}}))} \\
&    \times    \mathcal{L}_{I'_{\mathrm{D}} + I'_{\mathrm{U}}}(\mu y P_{\mathrm{U}}^{-1} K_{1}^{(\epsilon-1)} R^{\alpha_{1}(1-\epsilon )}),
\end{aligned}
\end{equation}
where the Laplace transform of $(I'_{\mathrm{D}} + I'_{\mathrm{U}})$, assuming $s=\mu y P_{\mathrm{U}}^{-1} K_{1}^{-1} R^{\alpha_{1}}$, is given by
\begin{equation}\label{eq:laplace_FD_U_1}
\begin{aligned}
 &\mathcal{L}_{I'_{\mathrm{D}} + I'_{\mathrm{U}}}(s) = \underbrace{\mathbb{E}_{\Phi_{\mathrm{B}}^{\mathrm{F}}       \cup       \Phi_{\mathrm{B}}^{\mathrm{D}}, g'_b} \Bigg[ e^{-s \sum_{b      \in \{\Phi_{\mathrm{B}}^{\mathrm{D}}       \cup       \Phi_{\mathrm{B}}^{\mathrm{F}} \backslash{b_0}\}}\: {g'_b P_{\mathrm{B}} K_{3}L_b^{-\alpha_{3}}}} \Bigg]}_{H_x(s)}    \times   \\
&\underbrace{\mathbb{E}_{\tilde{\Phi}_{\mathrm{U}}^{\mathrm{F,U}}      \cup       \tilde{\Phi}_{\mathrm{U}}^{\mathrm{H,U}}, h'_u, Z_u}\Bigg[e^{-s \sum_{u      \in \{\tilde{\Phi}_{\mathrm{U}}^{\mathrm{F,U}}       \cup       \tilde{\Phi}_{\mathrm{U}}^{\mathrm{H,U}}\}: X_u > Z_u }\: {\frac{h'_u P_{\mathrm{U}} Z_u^{\epsilon \alpha_1}}{K_{1}^{(\epsilon-1)} X_u^{\alpha_1}}}} \Bigg]}_{H_y(s)}.
\end{aligned}
\end{equation}

The first term in (\ref{eq:laplace_FD_U_1}) can be further written as,
%\begin{equation}\label{eq:laplace_FD_U_2}
%\mathrm{exp} \left( -2\pi \rho_{\mathrm{F}} \lambda_{\mathrm{B}} \int_{0}^{\infty} \left(\frac{s K_{3} P_{\mathrm{B}} v^{-\alpha_{3}}}{s K_{3} P_{\mathrm{B}} v^{-\alpha_{3}} + \mu } \right) v \mathrm{d}v \right)       \cdot \mathrm{exp} \left( -2\pi \rho_{\mathrm{D}} \lambda_{\mathrm{B}} \int_{0}^{\infty} \left(\frac{s K_{3} P_{\mathrm{B}} v^{-\alpha_{3}}}{s K_{3} P_{\mathrm{B}} v^{-\alpha_{3}} + \mu } \right) v \mathrm{d}v \right).
%\end{equation}

%Eq. (\ref{eq:laplace_FD_U_2}) can be simplified as:
\vspace{-3mm}
\begin{equation}\label{eq:laplace_FD_U_2A}
H_x(s) = e^{-2\pi (\rho_{\mathrm{F}} + \rho_{\mathrm{D}}) \lambda_{\mathrm{B}} \int_{0}^{\infty} \left(\frac{s K_{3} P_{\mathrm{B}} v^{-\alpha_{3}}}{s K_{3} P_{\mathrm{B}} v^{-\alpha_{3}} + \mu } \right) v \mathrm{d}v}.
\vspace{-1mm}
\end{equation}

The lower extreme of integration in the above term is zero because the closest interferer BS (either FD or HD) can be at any distance greater than zero. The second term can be further written as:
%\begin{equation}\label{eq:laplace_FD_U_3}
%\begin{aligned}
% \mathrm{exp} &\left( -2\pi \rho_{\mathrm{F}} \lambda_{\mathrm{B}} \int_{0}^{\infty}  \left(1 - \mathbb{E}_{Z_u}\left[\frac{\mu }{s P_{\mathrm{U}} K_1^{1-\epsilon } Z_u^{\epsilon \alpha_1} v^{-\alpha_{1}} \mathds{1}\{Z_u < v\} + \mu }\right] \right) v \mathrm{d}v \right)      \cdot \\
%& \mathrm{exp} \left( -2\pi \rho_{\mathrm{U}} \lambda_{\mathrm{B}} \int_{0}^{\infty} \left(1-\mathbb{E}_{Z_u}\left[\frac{\mu }{s P_{\mathrm{U}} K_1^{1-\epsilon } Z_u^{\epsilon \alpha_1} v^{-\alpha_{1}}  \mathds{1}\{Z_u < v\}  + \mu } \right] \right) v \mathrm{d}v \right).
%\end{aligned}
%\end{equation}
\vspace{-1mm}
\begin{equation}\label{eq:laplace_FD_U_3}
\begin{aligned}
&H_y(s) = \\
&e^{-2\pi (\rho_{\mathrm{F}} + \rho_{\mathrm{U}}) \lambda_{\mathrm{B}} \int_{0}^{\infty}  \left(1 - \mathbb{E}_{Z_u}\left[\frac{\mu }{s P_{\mathrm{U}} K_1^{1-\epsilon } Z_u^{\epsilon \alpha_1} v^{-\alpha_{1}} \mathds{1}\{Z_u < v\} + \mu }\right] \right) v \mathrm{d}v }.
\end{aligned}
\end{equation}

The lower extreme of integration in the above term is also zero but the constraint $\{Z_u < v\}$ makes sure that only active UEs from the other cells are included in the interference term. The terms in expression (\ref{eq:laplace_FD_U_3}) are further solved in Appendix~\ref{app:B}. Under the special case of no power control ($\epsilon = 0$), the above expression can be written as:
\begin{equation}\label{eq:laplace_FD_U_4A}
\begin{aligned}
H_y(s) = e^{ -2\pi  ( \rho_{\mathrm{F}} + \rho_{\mathrm{U}} ) \lambda_{\mathrm{B}} \int_{0}^{\infty}  \left( \frac{s P_{\mathrm{U}} K_1 v^{-\alpha_1}}{\mu + s P_{\mathrm{U}} K_1 v^{-\alpha_1}}\right) \mathbb{P} (Z_u \leq v) v \mathrm{d}v},
\end{aligned}
\end{equation}
where $\mathbb{P} (Z_u \leq v) $ is given in~(\ref{eq:distance_dis_cdf_mod}). Finally, by plugging~(\ref{eq:sinr_ccdf_u_2}) in (\ref{eq:sinr_ccdf_u_1}), we obtain the CCDF of the uplink SINR in a FD cell for a mixed system.
\begin{equation}\label{eq:sinr_ccdf_u_finally}
\begin{split}
&\mathrm{P}[\gamma_{\mathrm{FD,BS}} > y] =\\
 &\int_0^{\infty} e^{-\mu y P_{\mathrm{U}}^{-1} K_{1}^{(\epsilon-1)} R^{\alpha_{1}(1-\epsilon )} (N_{1}+\mathcal{C}(P_{\mathrm{B}}))} H_x(s) H_y(s) f_r(R) dR,
\end{split}
\end{equation}
where $s=\mu y P_{\mathrm{U}}^{-1} K_{1}^{-1} R^{\alpha_{1}}$, and $f_r(R)$ is given by~(\ref{eq:distance_dis_pdf_mod}); note that we will determine the value to be used for the parameter $\nu$ of be used for the parameter~(\ref{eq:distance_dis_pdf_mod}) in Appendix~\ref{app:A}, where we will compare the analytical model with simulation results.

\subsection{Downlink and Uplink SINR in the HD Cells of the Mixed System}\label{sec:HDdownlinkSINR}
The downlink SINR at a UE in a HD cell of the mixed system can be derived similarly to the downlink SINR in a FD cell. A downlink UE in a HD cell gets interference from all simultaneous uplink and downlink transmissions similar to the downlink UE in a FD cell. However, there will one difference from the derivation of the SINR CCDF given in Section~\ref{sec:FDdownlinkSINRCCDF}. To consider the interference from all the active uplink transmissions, the lower extreme of integration in (\ref{eq:laplace_FD_D_5}) is zero, which includes the uplink transmission in its own FD cell, whereas in the case of a HD cell, we need to make sure that no uplink transmission inside the downlink UE's own cell is included. For analytical tractability, to take this into account, we make an approximation that the distance from the nearest interfering uplink transmission is approximated by the distance from the nearest interfering BS. This is the same approximation made in \cite{Quekhybrid, elsawy_alpha_duplex} while modeling the UE-to-UE interference at a FD UE.

Thus, in this case, the lower extreme of  integration in (\ref{eq:laplace_FD_D_5}) will be $R$, i.e., the distance of the downlink UE from its serving BS. For this case,
\vspace{-1mm}
\begin{equation}\label{eq:laplace_FD_D_5_downlink_HD}
\begin{aligned}
&L'_y(s) =  \\
& e^{ -2\pi (\rho_{\mathrm{F}} + \rho_{\mathrm{U}}) \lambda_{\mathrm{B}} \int_{R}^{\infty}  \Bigg(1 - \mathbb{E}_{Z_u}\left[\frac{\mu }{s K_{2} P_{\mathrm{U}} K_1^{-\epsilon } Z_u^{\epsilon \alpha_1} v^{-\alpha_{2}} + \mu }\right] \Bigg) v \mathrm{d}v} .
\end{aligned}
\end{equation}

Similar to~(\ref{eq:sinr_downlink_finally}), the expression for CCDF of $\gamma_{\mathrm{HD,UE}}$, is given by,
\vspace{-1mm}
\begin{equation}\label{eq:sinr_hd_down_CCDF}
\begin{aligned}
&\mathrm{P}[\gamma_{\mathrm{HD,UE}} > y] = \\
& \int_0^{\infty} e^{-\mu y P_{\mathrm{B}}^{-1} K_{1}^{-1} R^{\alpha_{1}} N_{0}}~L_x(s)~L'_y(s) f_r(R) \mathrm{d}R,
\end{aligned}
\end{equation}
where $s = \mu y P_{\mathrm{B}}^{-1} K_{1}^{-1} R^{\alpha_{1}}$.

In the uplink case, the expression for uplink SINR in a HD cell will be given as the uplink SINR in a FD cell in Section~\ref{sec:FDuplinkSINR} but without any self-interference, i.e.,~$\mathcal{C}(P_{\mathrm{B}}) = 0$,
\vspace{-1mm}
\begin{equation}\label{eq:sinr_hd_up}
\gamma_{\mathrm{HD,BS}}=\frac{P_{\mathrm{RX,BS}}}{N_{1}+I'_{\mathrm{D}}+I'_{\mathrm{U}}}.
\end{equation}

The CCDF of $\gamma_{\mathrm{HD,BS}}$ is given by,
\begin{equation}\label{eq:sinr_ccdf_u_1_hd}
\mathrm{P}[\gamma_{\mathrm{HD,BS}} > y] = \int_0^{\infty} \mathrm{P}[\gamma_{\mathrm{HD,BS}} > y|r=R] f_r(R) \mathrm{d}R,
\end{equation}
\vspace{-1mm}
where
\begin{equation}\label{eq:sinr_ccdf_u_2_hd}
\begin{aligned}
&\mathrm{P}[\gamma_{\mathrm{HD,BS}} > y|r=R] = \\
&e^{-\mu y P_{\mathrm{U}}^{-1} K_{1}^{(\epsilon-1)} R^{\alpha_{1}(1-\epsilon )} N_{1}}~\mathcal{L}_{I'_{\mathrm{D}} + I'_{\mathrm{U}}}(\mu y P_{\mathrm{U}}^{-1} K_{1}^{(\epsilon-1)} R^{\alpha_{1}(1-\epsilon)}),
\end{aligned}
\end{equation}
where, for $s= \mu y P_{\mathrm{U}}^{-1} K_{1}^{-1} R^{\alpha_{1}}$, the expression for $\mathcal{L}_{I'_{\mathrm{D}} + I'_{\mathrm{U}}}(s)$ is same as given in (\ref{eq:laplace_FD_U_1}).

\section{Average rate}
In general, the average rate per hertz can be computed as follows \cite{jeffAndrews_uplink}.
\begin{align}
\mathbb{E}[\mathrm{C}] =~\mathbb{E}\left[\log_{2}(1+\gamma )\right]= \int
_{0}^{\infty}\mathrm{P}\left[\log_{2}(1+\gamma )>u\right]\mathrm{d}u.
\end{align}

By applying this, we can derive the average rate for the downlink and uplink in both FD and HD cells. By using~(\ref{eq:sinr_ccdf_general}), and (\ref{eq:sinr_ccdf_2}), the average downlink rate in the FD cell is given by
\begin{equation}\label{eq:rate_donwlink_FD}
\begin{aligned}
&\mathbb{E}[\mathrm{C_{FD,UE}}] = \int_{0}^{\infty}\mathrm{P}\left[\log_{2}(1+\gamma_{FD,UE})>u\right]\mathrm{d}u \\
&= \int _{0}^{\infty}\int_{0}^{\infty}  e^{-\mu (2^{u}-1) P_{\mathrm{B}}^{-1} K_{1}^{-1} R^{\alpha_{1}} N_{0}}    \times   \\
&\mathcal{L}_{I_{\mathrm{D}} + I_{\mathrm{U}}}(\mu (2^{u}-1) P_{\mathrm{B}}^{-1} K_{1}^{-1} R^{\alpha_{1}})f_{r}(R)~\mathrm{d}R~\mathrm{d}u.
\end{aligned}
\end{equation}

By using (\ref{eq:sinr_ccdf_u_1}), and (\ref{eq:sinr_ccdf_u_2}), the average uplink rate in a FD cell is given by
\begin{equation}\label{eq:rate_uplink_FD}
\begin{aligned}
&\mathbb{E}[\mathrm{C_{FD,BS}}] = \int_{0}^{\infty}\mathrm{P}\left[\log_{2}(1+\gamma_{FD,BS})>u\right]\mathrm{d}u \\
& = \int _{0}^{\infty}\int_{0}^{\infty}  e^{-\mu (2^{u}-1) P_{\mathrm{U}}^{-1} K_{1}^{(\epsilon-1)} R^{\alpha_{1}(1-\epsilon )} (N_{1}+\mathcal{C}(P_{\mathrm{B}}))}    \times    \\
&\mathcal{L}_{I'_{\mathrm{D}} + I'_{\mathrm{U}}}(\mu (2^{u}-1) P_{\mathrm{U}}^{-1} K_{1}^{-1} R^{\alpha_{1}})f_{r}(R)~\mathrm{d}R~\mathrm{d}u.
\end{aligned}
\end{equation}

Similarly the average downlink and uplink rates in a HD cell, i.e, $\mathrm{E}[\mathrm{C_{HD,UE}}]$, $\mathrm{E}[\mathrm{C_{HD,BS}}]$, respectively, can be derived.

Combining the rates of FD and HD cells, the average downlink and uplink rates of the complete network are given by, respectively,
\begin{equation}\label{eq:rate_donwlink_complete}
\mathbb{E}[\mathrm{C_{D}}] = \rho_F \mathbb{E}[\mathrm{C_{FD,UE}}] + \rho_D \mathbb{E}[\mathrm{C_{HD,UE}}]
\end{equation}
\begin{equation}\label{eq:rate_uplink_complete}
\mathbb{E}[\mathrm{C_{U}}] = \rho_F \mathbb{E}[\mathrm{C_{FD,BS}}] + \rho_U \mathbb{E}[\mathrm{C_{HD,BS}}]
\end{equation}

The ASE can be obtained as the product of the average spectral efficiency and of the density, the downlink and uplink ASEs of the mixed network can be obtained from~(\ref{eq:rate_donwlink_complete}) and~(\ref{eq:rate_uplink_complete}), respectively, as follows:
\begin{equation}\label{eq:ASE_donwlink}
\mathrm{ASE_D} = \lambda_{\mathrm{B}}\mathbb{E}[\mathrm{C_{D}}] = \lambda_{\mathrm{B}}(\rho_F \mathbb{E}[\mathrm{C_{FD,UE}}] + \rho_D \mathbb{E}[\mathrm{C_{HD,UE}}])
\end{equation}
\begin{equation}\label{eq:ASE_uplink}
\mathrm{ASE_U} = \lambda_{\mathrm{B}}\mathbb{E}[\mathrm{C_{U}}] = \lambda_{\mathrm{B}}(\rho_F \mathbb{E}[\mathrm{C_{FD,BS}}] + \rho_D \mathbb{E}[\mathrm{C_{HD,BS}}])
\end{equation}

\section{Numerical Results}
The formulation we presented in Section~\ref{sec:MainSINR} has been obtained using some approximations that allow us to keep the mathematical tractability of the model, so that to obtain the SINR CCDF for the downlink and uplink. Before being used, though, our model needs to be validated, in order to ensure that it can provide trustable results. We use some simulations to generate the expected SINR CDF curves for both the downlink and uplink, and to match the analytical curves. The model benchmark can be found in the Appendix~\ref{app:A}, which the reader can refer to for further details.

This benchmark serves also as tool to determine what value of $\nu$ should be used for the analytical model (see Section~\ref{sec:Th_ANA_pdf_dis}). It turns out that, in the downlink, $\nu = 1$ gives a better match, while $\nu = 1.25$ provides better results for the uplink case. Therefore, in the following sections, we will use $\nu = 1$ and $\nu = 1.25$ to compute the numerical results for the downlink and for the uplink, respectively.

We evaluate the throughput of the proposed mixed system, and present the effect on it of network parameters such as SIC, and the transmit powers of BS and UEs. The performance of the mixed system is also compared with a traditional synchronous TDD half duplex system (THD System), in which, (1) in a given time slot, all cells schedule either uplink or downlink transmission, and (2) the number of time slots is divided equally between the uplink and downlink transmission. In this case, a downlink transmission receives interference from only the neighboring BSs and an uplink transmission receives interference from only the uplink transmissions of the neighboring cells.

\begin {table}
\caption {Network Parameters} \label{tab:simulation_parameters}
\vspace{-3 mm}
\begin{center}
    \begin{tabular}{| p{1.5 in} | p{1.2 in} |}
    	\hline
		\textbf {Parameter} & \textbf{Value} \\ \hline
		Bandwidth & $10$ MHz \\ \hline
		BS Density [nodes/m$^2$]  & $10^{-3}$  \\ \hline
		Thermal Noise Density & $-174$ dBm/Hz \\ \hline
		Noise Figure & $9$ dB (UE), $8$ dB (BS)\\ \hline
		Path Loss (dB) ($R$ in km) \cite{3GPP36814} &  $140.7 + 36.7~\mathrm{log}_{10}(R)$ \\ \hline
		Outage SINR Threshold & $-8$ dB \\ \hline
    \end{tabular}
    \vspace{-7mm}
\end{center}
\end{table}

Please note that in our analysis in Section \ref{sec:MainSINR} we considered a general model where all the links have different channel parameters and the uplink has power control, however, in this section, we generate results for the specific case of using the same channel parameters for all the different links. The effect of different channel parameters for different links is left as future work. Moreover, we also assume that all uplink UEs transmit with the same power ($P_{\mathrm{U}}$), i.e., $\epsilon = 0$. We observed that power control modeled as in Section~\ref{sec:Powalloc} with $\epsilon \neq 0$ considerably lowers the uplink performance. This is due to the interference generated by the BSs, which do not implement any downlink power control. In HD networks with uplink power control, the UEs close to the BS reduce
their transmit power and so does the interference on other cells' uplink UEs; as a result, the cell-edge UEs' SINR improve. In contrast, in a FD network, even though the UEs close to the BS reduce their transmit power, the interference of the BS remain unchanged and, therefore, the UEs' SINR drops considerably, because the reduction of the received power is not compensated by a corresponding reduction of the interference. We reckon that an appropriate power control in FD networks is a challenge that should be addressed, and it will be considered in our future work.

\begin{figure}
\centering
\includegraphics[width = 2.6 in] {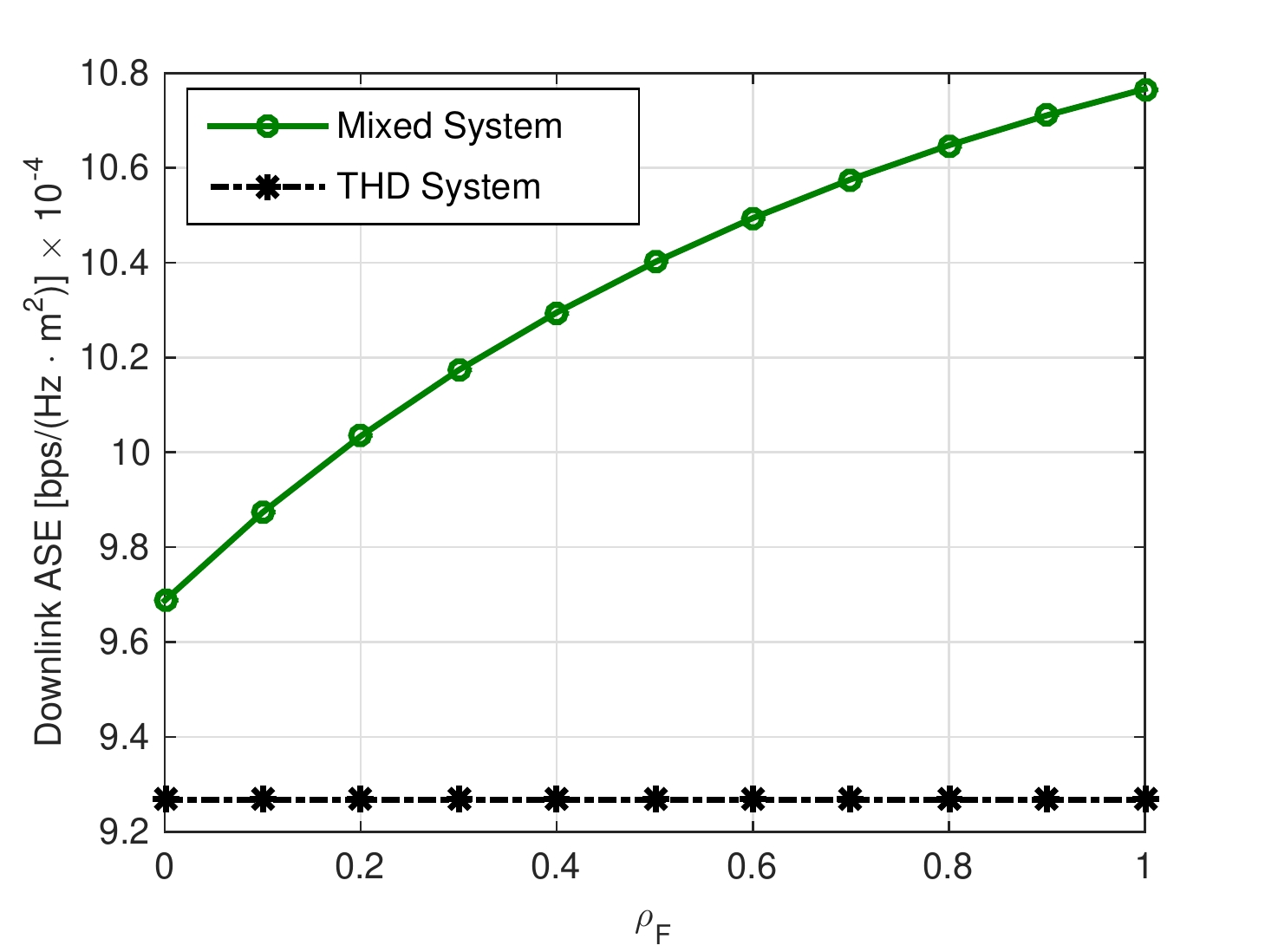}
\vspace{-2mm}
\caption{Downlink ASE as a function of the proportion of FD BSs ($\rho_{\mathrm{F}}$), where $\rho_{\mathrm{D}} = \rho_{\mathrm{U}} = (1 - \rho_{\mathrm{F}})/2$. The transmit powers, $P_{\mathrm{B}}$ = 24 dBm, $P_{\mathrm{U}}$ = 23 dBm. In THD system, $\rho_{\mathrm{D}} = 1$. }
\label{fig:dwn_ase_sic}
\end{figure}

\begin{figure}
\centering
\includegraphics[width = 2.6 in] {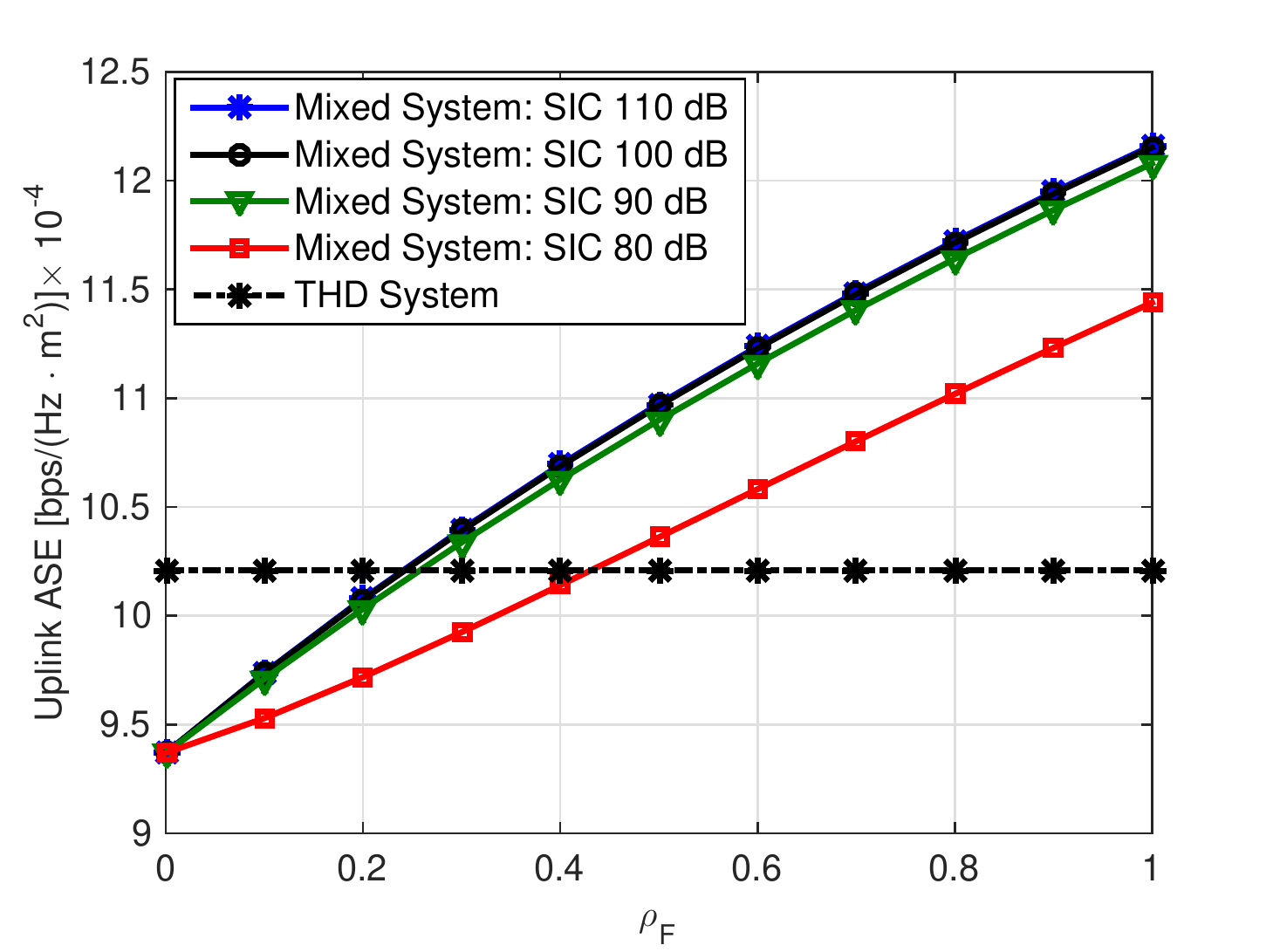}
\vspace{-2mm}
\caption{Uplink ASE as a function of the proportion of FD BSs ($\rho_{\mathrm{F}}$), where $\rho_{\mathrm{D}} = \rho_{\mathrm{U}} = (1 - \rho_{\mathrm{F}})/2$. The transmit powers, $P_{\mathrm{B}}$ = 24 dBm, $P_{\mathrm{U}}$ = 23 dBm. In THD system, $\rho_{\mathrm{U}} = 1$, $\mathcal{C}(P_{\mathrm{B}}) = 0$.}
\label{fig:up_ase_sic}
\end{figure}

We simulate a dense small cell network, for which the network parameter values are described in Table~\ref{tab:simulation_parameters}. With this setting we generate the following numerical results. Figs.~\ref{fig:dwn_ase_sic} and~\ref{fig:up_ase_sic} show the ASE of the mixed system as a function of the percentage of FD BSs ($\rho_{\mathrm{F}}$) with different SIC. The remaining BSs are equally divided into HD downlink and HD uplink modes, i.e., $\rho_{\mathrm{D}} = \rho_{\mathrm{U}} = (1 - \rho_{\mathrm{F}})/2$. The transmit power of the BS and the UE are fixed to 24 dBm, and 23 dBm, respectively.

\begin{figure}
\centering
\includegraphics[width = 2.6 in] {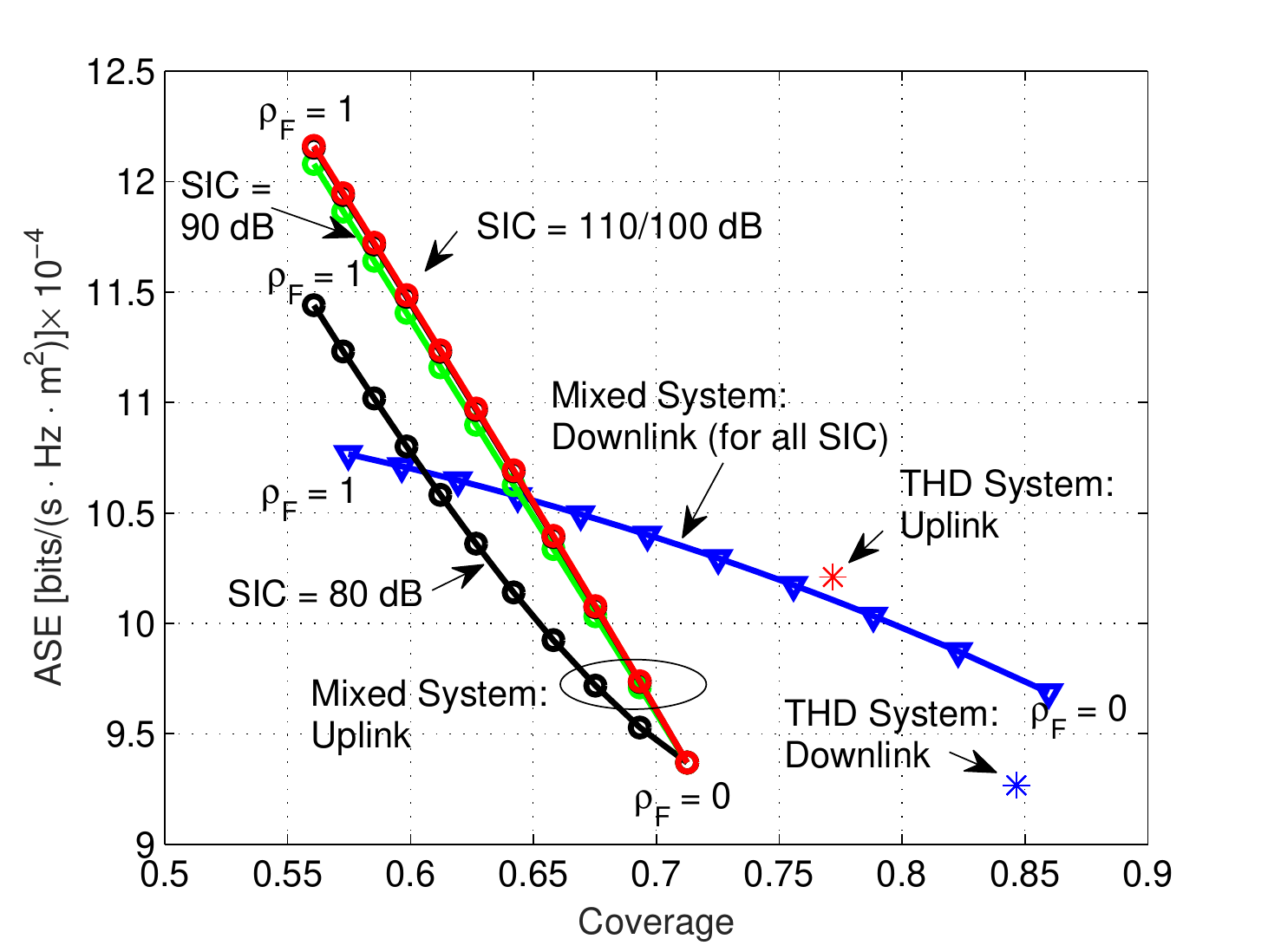}
\caption{ASE vs. Coverage, $P_{\mathrm{B}}$ = 24 dBm, $P_{\mathrm{U}}$ = 23 dBm.  For the mixed system, in the downlink, given the coverage of FD cells  $\theta_{\mathrm{FD,DL}}$ and of the HD cells $\theta_{\mathrm{HD,DL}}$, the overall downlink coverage of the mixed system is computed as $(\rho_{\mathrm{F}}\theta_{\mathrm{FD,DL}}+\rho_{\mathrm{D}} \theta_{\mathrm{HD,DL}})/(\rho_{\mathrm{F}}+\rho_{\mathrm{D}})$; similarly, the uplink coverage is obtained as $(\rho_{\mathrm{F}}\theta_{\mathrm{FD,UL}}+\rho_{\mathrm{U}} \theta_{\mathrm{HD,UL}})/(\rho_{\mathrm{F}}+\rho_{\mathrm{U}})$.}
\label{fig:coverage_ase_tradeoff}
\end{figure}

In the mixed system, as we increase the number of BSs in FD mode, both downlink and uplink ASE increase. As we increase $\rho_{\mathrm{F}}$, both the number of transmissions and the aggregated interference in each direction increase, which generates a tradeoff between ASE and the coverage as shown in Fig.~\ref{fig:coverage_ase_tradeoff}. We define coverage as the fraction of UEs in a non-outage region, where an outage happens if the received SINR goes below the outage SINR threshold. The increasing number of transmissions provides higher ASE as shown in Figs.~\ref{fig:dwn_ase_sic} and~\ref{fig:up_ase_sic}, but a higher outage as well as shown in Fig.~\ref{fig:coverage_ase_tradeoff}. The higher ASE gain is therefore achieved at the cost of lower coverage. Thus an appropriate ratio of FD BSs, reflecting a desired optimal tradeoff between these two conflicting objectives, should be enabled in the network.

As shown in Fig.~\ref{fig:dwn_ase_sic}, the throughput of the downlink direction is not affected by the SIC, because self-interference is received only in the uplink transmission. In the uplink direction, as shown in Fig.~\ref{fig:up_ase_sic}, the gain of the mixed system increases as SIC improves. It can also noted that there is no improvement in the uplink performance after reducing SIC below 100 dB, because for this dense multi small cell network, after some point inter-cell interference starts dominating the total interference in the uplink direction.

Moreover, Fig.~\ref{fig:up_ase_sic} shows that in the uplink direction, the mixed system is superior to THD system when the percentage of BSs in FD mode is higher than 25\%-45\%, depending on the SIC value, which is not the case in the downlink direction. For lower values of $\rho_{\mathrm{F}}$, in the mixed system most of the cells are in HD mode, similar to the THD system. However, in the mixed system the uplink transmission receives BS-to-BS interference from the cells, which are in HD downlink mode. This interference is generally stronger than the UE-to-BS interference, which decreases the throughput of the mixed uplink system compared to the uplink of THD system, where only UE-to-BS interference exists. However, in the downlink case, UE-to-UE interference is generally weaker than the BS-to-UE interference, which consequently leads to higher throughput in the mixed system even for lower values of $\rho_{\mathrm{F}}$.

\begin{figure}
\centering
\includegraphics[width = 2.6 in] {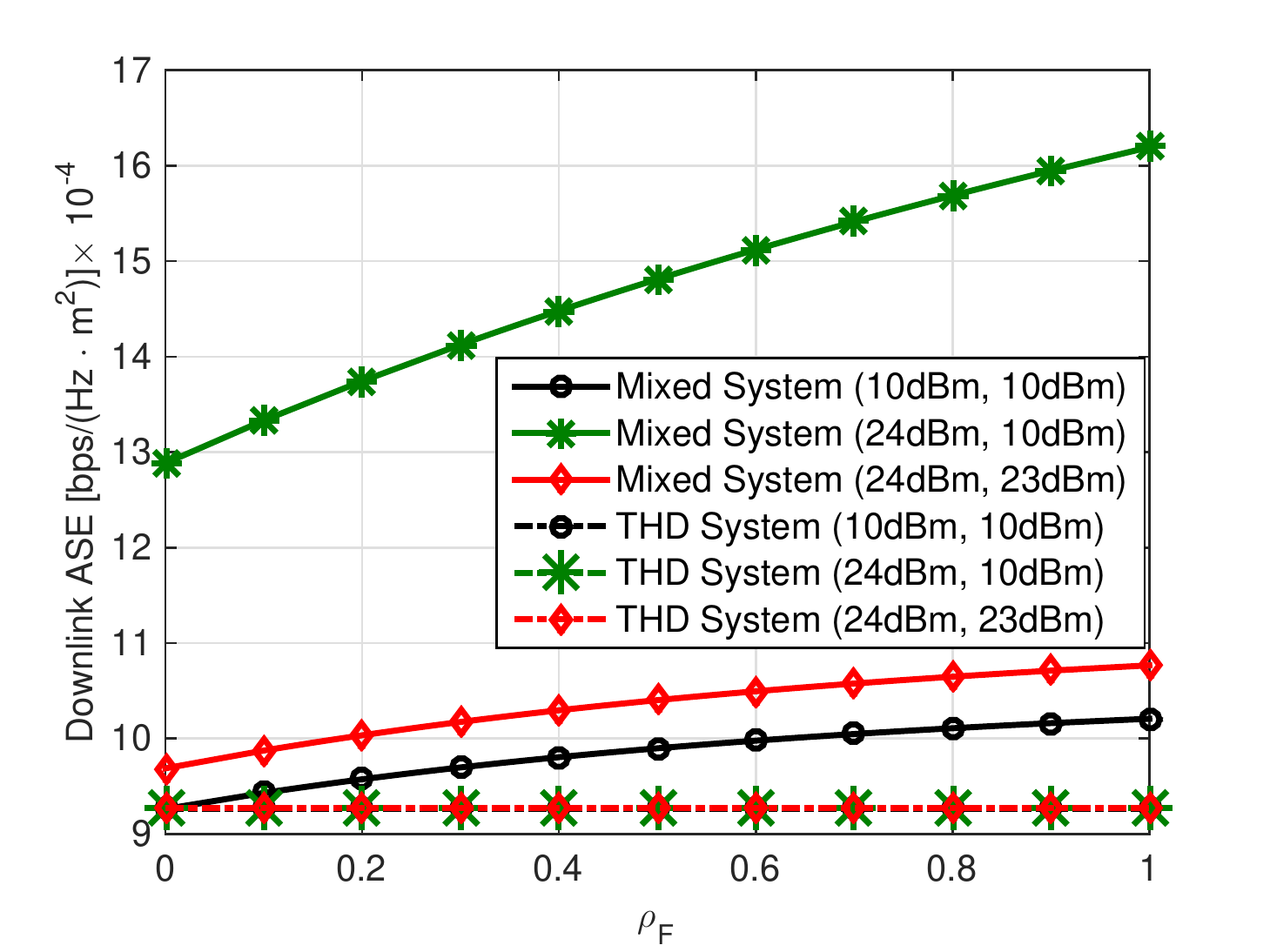}
\caption{Downlink ASE as a function of the proportion of FD BSs ($\rho_{\mathrm{F}}$) and with different BS and UE transmit powers ($P_{\mathrm{B}}, P_{\mathrm{U}}$). The other parameters: SIC = 110 dB, $\rho_{\mathrm{D}} = \rho_{\mathrm{U}} = (1 - \rho_{\mathrm{F}})/2$, for THD system, $\rho_{\mathrm{D}} = 1$. Note that all three THD plots overlap completely.}
\label{fig:dwn_ase_power}
\end{figure}

\begin{figure}
\centering
\includegraphics[width = 2.6 in] {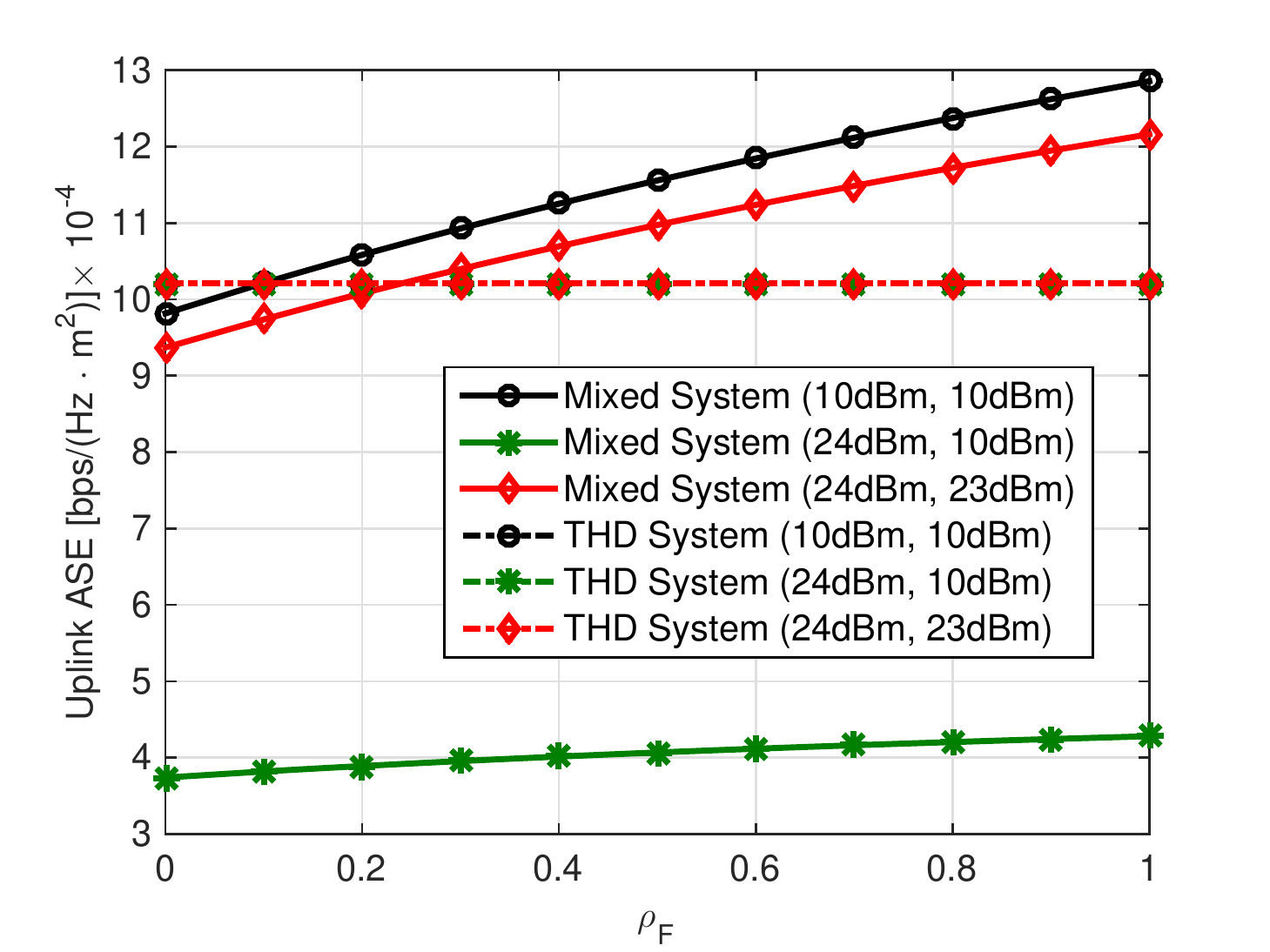}
\caption{Uplink ASE as a function of the proportion of FD BSs ($\rho_{\mathrm{F}}$) and with different BS and UE transmit powers ($P_{\mathrm{B}}, P_{\mathrm{U}}$). The other parameters: SIC = 110 dB, $\rho_{\mathrm{D}} = \rho_{\mathrm{U}} = (1 - \rho_{\mathrm{F}})/2$, for THD system, $\rho_{\mathrm{U}} = 1$, $\mathcal{C}(P_{\mathrm{B}}) = 0$. Note that all three THD plots overlap completely.}
\label{fig:up_ase_power}
\end{figure}

Figs.~\ref{fig:dwn_ase_power} and~\ref{fig:up_ase_power} show the impact of the transmit power of BS ($P_{\mathrm{B}}$) and UE ($P_{\mathrm{U}}$) on the downlink and uplink ASE. In the THD system, downlink performance depends only on $P_{\mathrm{B}}$ because the downlink transmission receives interference only from the neighboring downlink transmissions. Similarly, the uplink performance depends only on $P_{\mathrm{U}}$. In the THD system, changing the transmit power does not show much variation in any direction. This is because due to the high density of the BSs, it is an interference limited regime, and changing the transmit power in any direction also proportionately changes the interference, so the SINR does not vary much.

In the mixed system, both downlink and uplink performance depend on the transmit powers of both BS and UE. For the downlink case, as shown in Fig.~\ref{fig:dwn_ase_power}, as we reduce the uplink transmit power, it reduces the UE-to-UE interference, which improves the downlink throughput. The highest downlink gain in all the computed set of transmit powers is achieved when the difference between the downlink and uplink transmit power is maximum, which is the case with $P_{\mathrm{B}}$ = 24 dBm, and  $P_{\mathrm{U}}$ = 10 dBm in Fig.~\ref{fig:dwn_ase_power}. By contrast, in the mixed uplink case, as shown in Fig.~\ref{fig:up_ase_power}, the uplink gain improves as the difference between the downlink power and uplink power decreases. For example, in Fig.~\ref{fig:up_ase_power}, the highest uplink gain is achieved when the $P_{\mathrm{B}}$ is at the same level as $P_{\mathrm{U}}$.

\begin{figure}
\centering
\includegraphics[width = 2.7 in] {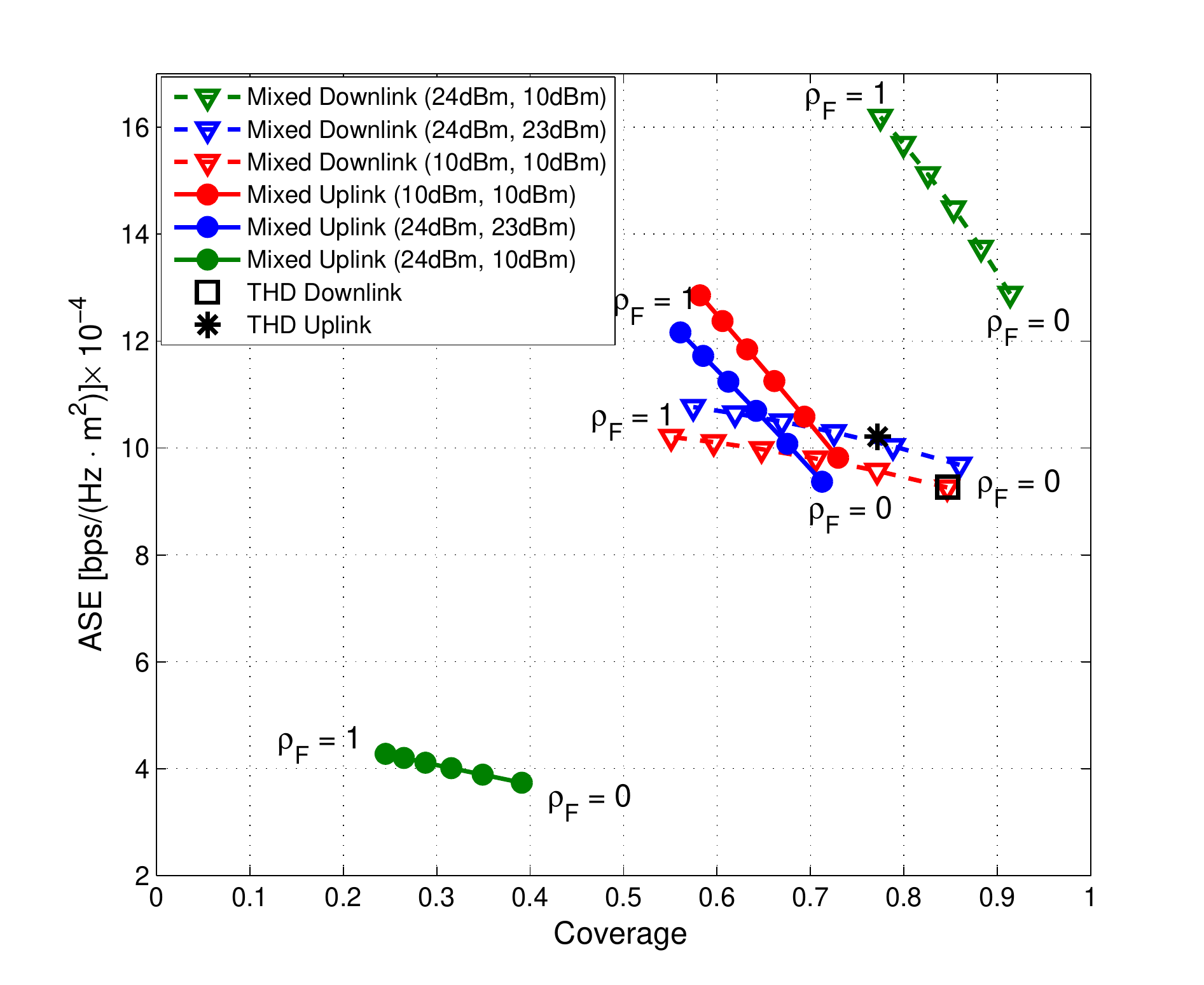}
\caption{ASE vs. Coverage with different BS and UE transmit powers ($P_{\mathrm{B}}, P_{\mathrm{U}}$), SIC = 110 dB.}
\label{fig:ase_coverage_powers}
\end{figure}

Fig.~\ref{fig:ase_coverage_powers} shows the tradeoff between ASE and coverage for a different set of transmit powers in downlink and uplink. The case of $P_{\mathrm{B}}$ = 24 dBm, and  $P_{\mathrm{U}}$ = 10 dBm provides the highest downlink coverage and downlink ASE but the worst uplink coverage and uplink ASE. These results show the need of an appropriate selection of transmit powers for joint performance gain in the mixed system. In general, to achieve the maximum joint uplink and downlink gain, both uplink and downlink powers should be optimized considering downlink and uplink UEs, as well as a pool of parameters, such as the BS-to-BS, BS-to-UE and UE-to-UE channels, SIC, etc. \cite{Goyal_arxiv}. In this paper, where we derive average analytical performance using fixed power allocation for all UEs, having similar transmit powers for BS and UE provides a fair performance to both uplink and downlink, while having unbalanced transmit powers benefits one direction at the cost of the other direction which may be a desirable outcome if the uplink and downlink traffic is asymmetric.

\section{Conclusion and future work}
In this paper we considered a mixed multi-cell system, composed of full duplex and half duplex cells, for which we proposed a stochastic geometry-based model that allows us to numerically assess the SINR complementary CDF and the average spectral efficiency, for both the downlink and uplink directions. Using this model, we studied the impact of FD cells on the average spectral efficiency vs. coverage tradeoff of these systems, for various transmit power values at the BS, at the UE, and for different self-interference cancellation levels.

We have shown that increasing the proportion of FD cells increases ASE but reduces coverage and, therefore, can be used as a design parameter of the network to achieve either a better ASE at the cost of limited coverage or a lower ASE with improved coverage, depending on the desired tradeoff between these two performance metrics. Moreover, we show that, in order for the downlink and uplink to achieve similar performance, the transmit power at the BS and at the UE should have similar values, but also that these powers can be tuned to achieve asymmetric uplink and downlink performance improvements if traffic demands dictate this. As future work, we will extend our study to include power control, and to different path loss models for the BS-to-BS, BS-to-UE and UE-to-UE channels.

\begin{appendices}

\section{}\label{app:A}
In Section~\ref{sec:MainSINR} we proposed a model to compute the downlink and uplink SINR CDF of a FD system in a multi-cell scenario. Due to the approximations we introduced to maintain the
mathematical tractability, a benchmark of the model is required in order to prove the accuracy of the proposed formulation. In particular, we compared results obtained by numerical integration of~(\ref{eq:sinr_downlink_finally}) and~(\ref{eq:sinr_ccdf_u_finally}) with those obtained through simulation. We neglect the self-interference and the noise and we assume that both UEs and BSs transmit with the same power. 
One of the goals of this benchmark is also to determine which value of $\nu$ should be used for the PDF of the distance to the transmitter (see Section~\ref{sec:Th_ANA_pdf_dis}); we evaluate two values, namely $\nu$ = 1 and $\nu$ = 1.25.

We resort to $\nu$ as a correction factor to compensate the lack of correlation among the process $\Phi_{\mathrm{B}}$ of BSs and $\tilde{\Phi}_{\mathrm{U}}$ of the active UEs, which we assumed to be independent in order to keep the analytical tractability of the model proposed in Section~\ref{sec:MainSINR}. Nonetheless, equation~(\ref{eq:distance_dis_pdf_gen}), which models the PDF of the distance to the closest point for SPPPs, does not reproduce the PDF of the distance to the closest point of the actual model, which is not an SPPP. The parameter $\nu$ allows us to adjust~(\ref{eq:distance_dis_pdf_gen}) to improve its match with the actual PDF obtained from the simulation results.
\begin{figure}
\centering
\includegraphics[width = 2.7 in] {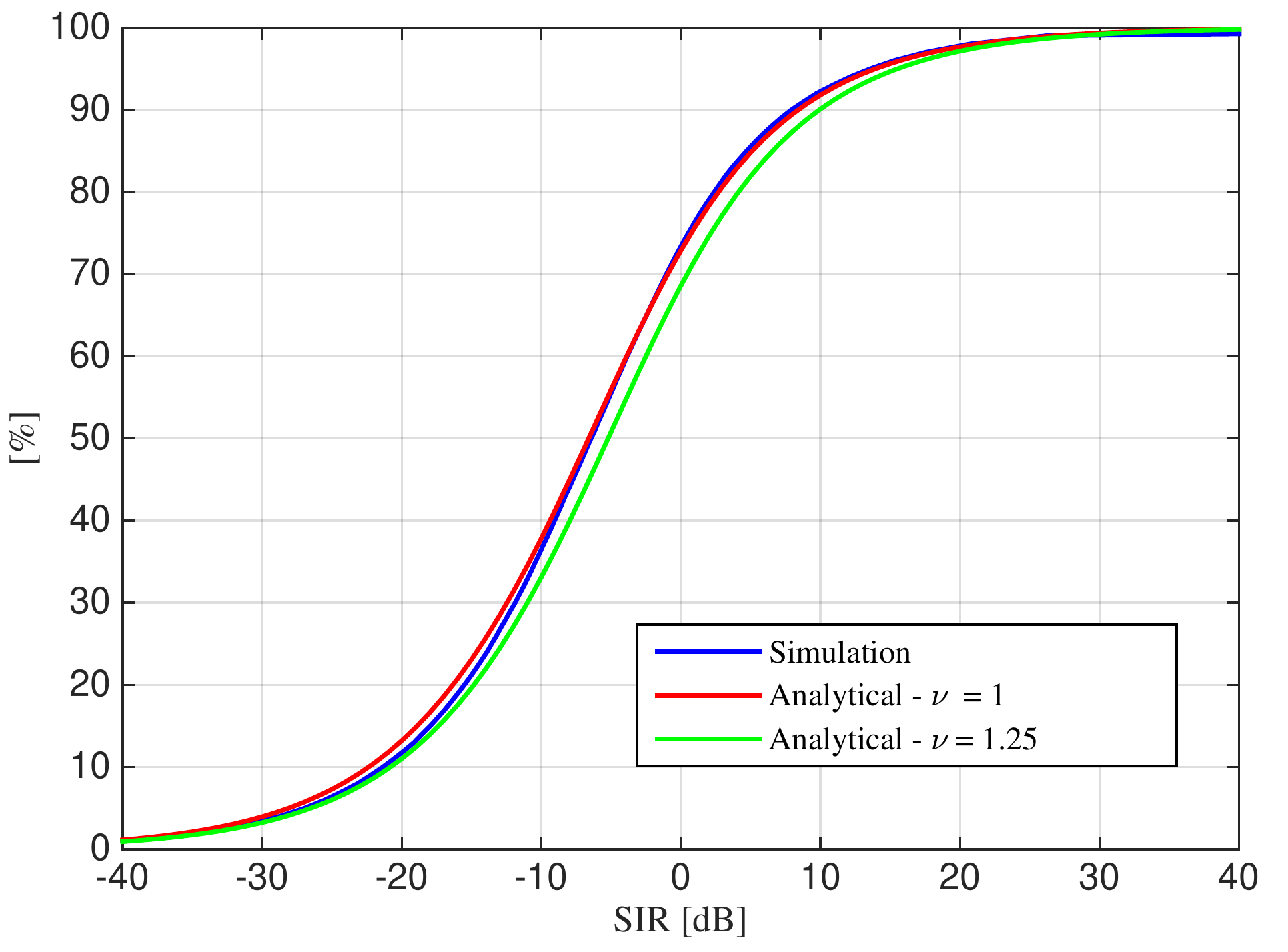}
\caption{Benchmark of the analytical model for the downlink SIR in mixed system.}
\label{fig:dwn_benchmark}
\end{figure}
\begin{figure}
\centering
\includegraphics[width = 2.7 in] {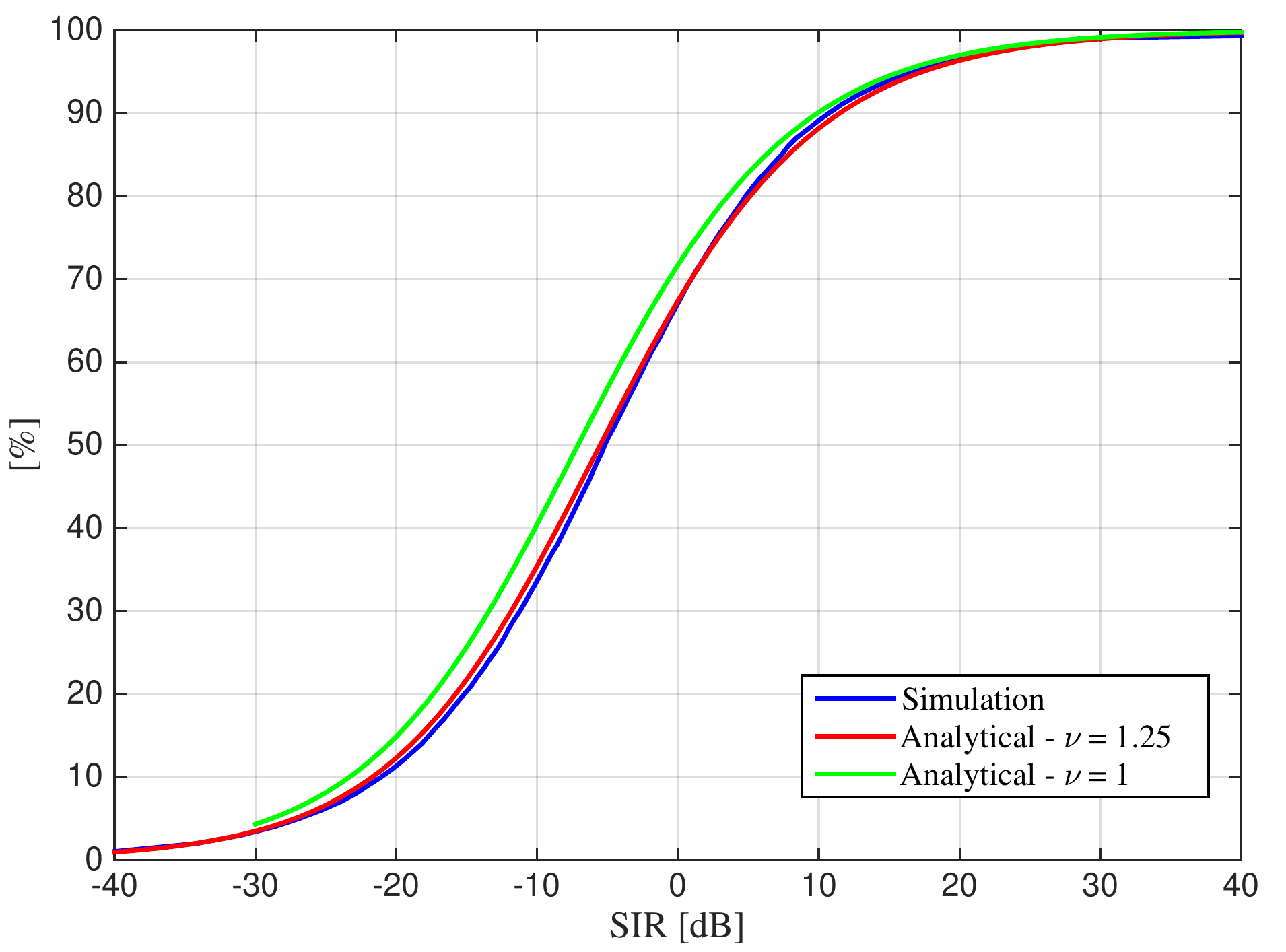}
\caption{Benchmark of the analytical model for the uplink SIR in mixed system.}
\label{fig:up_benchmark}
\end{figure}

We show the match of the analytical model for the downlink and uplink with the simulation results in Figs.~\ref{fig:dwn_benchmark} and ~\ref{fig:up_benchmark}. We can notice that $\nu = 1$ provides the best match for the downlink, while $\nu = 1.25$ gives the best results for the uplink. Therefore, we will use the value $\nu = 1$ to compute the numerical results for the downlink, whereas we will use $\nu = 1.25$ for the uplink.

\section{}\label{app:B}
\vspace{-1mm}
The integration in (\ref{eq:laplace_FD_U_3}) can be further solved as
\vspace{-1mm}
\begin{equation}\label{eq:laplace_FD_U_3_App_1}
\begin{aligned}
&\int_{0}^{\infty}  \left(1 - \mathbb{E}_{Z_u}\left[\frac{\mu }{s P_{\mathrm{U}} K_1^{1-\epsilon } Z_u^{\epsilon \alpha_1} v^{-\alpha_{1}} \mathds{1}\{Z_u < v\} + \mu }\right] \right) v \mathrm{d}v \\
&=\int_{0}^{\infty} v~ \mathbb{E}_{Z_u} \left(\frac{1}{\frac{\mu s^{-1} P_{\mathrm{U}}^{-1} K_1^{\epsilon - 1} Z_u^{-\epsilon \alpha_1} v^{\alpha_{1}} }{\mathds{1}\{Z_u < v\} } + 1} \right)  \mathrm{d}v  \\
%&= \int_{0}^{\infty} v~ \int_{0}^{v} \left(\frac{1}{\mu s^{-1} P_{\mathrm{U}}^{-1} K_1^{\epsilon - 1} z^{-\epsilon \alpha_1} v^{\alpha_{1}} + 1} \right) f_{Z_u}(z)~\mathrm{d}z~\mathrm{d}v  \\
&= \int_{0}^{\infty} v~ \int_{0}^{v} \left(\frac{1}{A~K_1^{\epsilon - 1}~z^{-\epsilon \alpha_1} v^{\alpha_{1}} + 1} \right) f_{Z_u}(z)~\mathrm{d}z~\mathrm{d}v  \\
& = G(A,K_1, \epsilon, \alpha_1)
\end{aligned}
\end{equation} where $A = \mu s^{-1} P_{\mathrm{U}}^{-1} $. By using integration by parts,  $G(A,K_1, \epsilon, \alpha_1) $ can be written as

%\begin{equation}\label{eq:laplace_FD_U_3_App_2}
%\begin{aligned}
%&=  \int_{0}^{\infty} v \left(\frac{1}{A~K_1^{\epsilon - 1} ~z^{-\epsilon \alpha_1} v^{\alpha_{1}} + 1} \right) \mathbb{P}\{ Z_u \leq z\}~\mathrm{d}v  \\
%& -  \int_{0}^{\infty} v \int_{0}^{v}  \frac{A~K_1^{\epsilon - 1}~\epsilon \alpha_1~z^{-(\epsilon \alpha_1 + 1)} v^{\alpha_{1}}}{({A~K_1^{\epsilon - 1}~z^{-\epsilon \alpha_1} v^{\alpha_{1}} + 1})^2}  \mathbb{P}\{ Z_u \leq z\}~\mathrm{d}z~\mathrm{d}v
%\end{aligned}
%\end{equation}
\begin{equation*}
=  \int_{0}^{\infty} v \left(\frac{1}{A~K_1^{\epsilon - 1} ~z^{-\epsilon \alpha_1} v^{\alpha_{1}} + 1} \right) \mathbb{P}\{ Z_u \leq z\}~\mathrm{d}v  
\end{equation*}
\begin{equation}\label{eq:laplace_FD_U_3_App_2}
 -  \int_{0}^{\infty} v \int_{0}^{v}  \frac{A~K_1^{\epsilon - 1}~\epsilon \alpha_1~z^{-(\epsilon \alpha_1 + 1)} v^{\alpha_{1}}}{({A~K_1^{\epsilon - 1}~z^{-\epsilon \alpha_1} v^{\alpha_{1}} + 1})^2}  \mathbb{P}\{ Z_u \leq z\}~\mathrm{d}z~\mathrm{d}v
\end{equation}
\vspace{-4mm}
\begin{equation}\label{eq:laplace_FD_U_3_App_3}
\begin{aligned}
& \text{For $\epsilon = 0$,} ~ G(A,K_1, 0, \alpha_1)   \\
&=\int_{0}^{\infty} v \left(\frac{1}{A~K_1^{- 1} z^{-\epsilon \alpha_1} v^{\alpha_{1}} + 1} \right) \mathbb{P}\{ Z_u \leq z\}~\mathrm{d}v
\end{aligned}
\end{equation}
\end{appendices}
\vspace{-2mm}
\bibliographystyle{IEEEtran}
\bibliography{FD_references}

\end{document}